\documentclass[twocolumn,journal]{IEEEtran}

\usepackage{amsfonts}
\usepackage{amsmath}
\usepackage{amsthm}
\usepackage{amssymb}
\usepackage{graphicx}
\usepackage[T1]{fontenc}
\usepackage{supertabular}
\usepackage{longtable}
\usepackage[usenames,dvipsnames]{color}
\usepackage{bbm}
\usepackage{fancyhdr}
\usepackage{breqn}
\usepackage{fixltx2e}
\usepackage{capt-of}
\usepackage{tikz}
\usetikzlibrary{matrix}
\usepackage{endnotes}
\usepackage{soul}
\usepackage{marginnote}

\usepackage{multirow}
\newcommand{\mathsym}[1]{}
\newcommand{\unicode}[1]{}

\usepackage{hyperref}

\usepackage{graphicx}
\graphicspath{{figures/}{/Users/nntaleb/Dropbox/Latex/SilentRisk/}}
 
\usepackage[framemethod=TikZ]{mdframed}
\usepackage[framemethod=TikZ]{mdframed}
\usepackage{framed}

\usepackage{enumitem}

\graphicspath{{figures/}{/Users/nntaleb/Dropbox/Latex/Collectednew/}}
\mdfsetup{%
nobreak=false,
   middlelinecolor=black,
   middlelinewidth=1pt,
   backgroundcolor=yellow!20,
   roundcorner=3pt}

\mdtheorem[nobreak=true,outerlinewidth=1,
backgroundcolor=yellow!50, outerlinecolor=black,innertopmargin=0pt,splittopskip=\topskip,skipbelow=\baselineskip, skipabove=\baselineskip,ntheorem,roundcorner=5pt,font=\itshape]{theorem}{Theorem}

\mdtheorem[nobreak=true,outerlinewidth=1,
backgroundcolor=yellow!20, outerlinecolor=black,innertopmargin=0pt,splittopskip=\topskip,skipbelow=\baselineskip, skipabove=\baselineskip,ntheorem,roundcorner=5pt,font=\itshape]{remark}{Comment}

\mdtheorem[nobreak=true,outerlinewidth=1,
backgroundcolor=pink!30, outerlinecolor=black,innertopmargin=0pt,splittopskip=\topskip,skipbelow=\baselineskip, skipabove=\baselineskip,ntheorem,roundcorner=5pt,font=\itshape]{principle}{Principle}

\mdtheorem[nobreak=true,outerlinewidth=1,
backgroundcolor=yellow!50, outerlinecolor=black,innertopmargin=5pt,splittopskip=\topskip,skipbelow=\baselineskip, skipabove=\baselineskip,ntheorem,roundcorner=5pt,font=\itshape]{inbrief}{In Brief}

\usepackage{hyperref}

\title{\color{Brown}Hidden Risks and Optionalities in American Options}

\author{
\IEEEauthorblockN{
Noura El Hassan\IEEEauthorrefmark{1},
Bacel Maddah\IEEEauthorrefmark{2},
Nassim Nicholas Taleb\IEEEauthorrefmark{2}\IEEEauthorrefmark{3}
}
\\
\IEEEauthorblockA{\IEEEauthorrefmark{1}
American University of Beirut--Mediterraneo\\
Paphos, Cyprus}
\\
\IEEEauthorblockA{\IEEEauthorrefmark{2}
American University of Beirut, Beirut, Lebanon}
\\
\IEEEauthorblockA{\IEEEauthorrefmark{3}
Universa Investments\\
nnt@fooledbyrandomness.com}
}

\begin{document}
\maketitle
\begin{mdframed}
	\begin{abstract}
		We develop a practical framework for identifying and quantifying the hidden layers of risks and optionality embedded in American options by introducing stochasticity into one or more of their underlying determinants. The heuristic approach remedies the problems of conventional pricing systems, which treat some key inputs deterministically, hence systematically underestimate the flexibility and convexity inherent in early-exercise features.
	\end{abstract}
\end{mdframed}

\section{Unaccounted Optionality}\label{sec:intro}

\subsection{A note on model nonlinearity as fragility}\label{fragility}
Fragility to model error has been mapped in terms of convexity \cite{taleb2013mathematical}, and its heuristic testing presented below applied among others by the IMF to gauge portfolio risks of banks, see \cite{taleb2018IMF}. 

The logic is as follows. Having the right model, but being subjected to parameter uncertainty will invariably lead to an expected increase in model error in the presence of convexity and nonlinearities, particularly when the second order effect is not negligible. Assume $f(.)$ an estimated function:
\begin{equation}
  f(x \mid a), 
\end{equation}
where $a$ is fixed, assumed to be the average or expected parameter, taking $u$ as the distribution of $a$ over its domain $\mathcal{A}$:
\begin{equation}\label{eq:10.6-amean}
  \bar a \;=\; \int_{\mathcal{A}} a\,u(a)\,da. \tag{6}
\end{equation}
The mere fact that $a$ is uncertain (since it is estimated) might lead to a bias if we perturbate from the outside (of the integral), i.e.\ stochasticize the parameter deemed fixed. Accordingly, fragility to model error $\pi_A$ is easily measured as the difference between (a) $f$ integrated across values of potential $a$ and (b) $f$ estimated for a single value of $a$ deemed to be its average. The convexity bias $\pi$ becomes

\begin{multline}
  \pi_A \;=\;
  \int_{\mathcal{X}} \int_{\mathcal{A}} f(x \mid a)\,u(a)\,da\,dx
  \;-\\  
  \int_{\mathcal{X}} f\!\left(x \,\middle|\, \int_{\mathcal{A}} a\,u(a)\,da \right)\!dx. \label{eqforfragility}
\end{multline}

This can be approximated by an interpolated estimate obtained with two values of $a$ separated from a mid-point by $\Delta a$, the mean deviation of $a$, and estimating the convexity bias $\pi_{\Delta a}$

\begin{multline}
	\pi_{\Delta a} \;\approx\;
  \int_{-\infty}^{K} \frac{1}{2}\Big(f(x \mid \bar a+\Delta a)\\
  +f(x \mid \bar a-\Delta a)\Big)\,dx
  \;-\;
  \int_{-\infty}^{K} f(x \mid \bar a)\,dx.\label{simplifiedeq}
\end{multline}
  
Furthermore, particularly in the case of options, even if a pricing approximation is used, the result may not illuminate us on option valuation but will give us a degree of model risk. Under the principle in \cite{taleb2013mathematical}, a bad ruler might not give us the precise height of a growing child, but will inform us whether the child is growing. As we are looking for fragilities, this allows us some approximations that work well with otherwise computationally onerous American options.

\subsection{Application to American Options}
American options differ from their European counterparts in allowing early exercise. This single feature introduces a set of nonlinearities and latent exposures that remain largely invisible to conventional risk systems. 

A standard option $O(.)$ is a function
\begin{equation}
O(S, K, \sigma, T, r_1, r_2),	\label{initial}
\end{equation}
where $S$ is the underlying security price at time 0, $K$ the strike price, $\sigma$ the volatility, $T$ the time to nominal expiration, $r_1$ the funding rate, and $r_2$ the ``carry'' of the underlying (which could be the discrete dividend or continuous foreign rate).

Under conventional pricing models (starting with \cite{Bachelier1900} ), only $S$ is stochastic. In further refinements and adaptations, $\sigma$ is treated as stochastic, with a rich literature \cite{Dupire1994, Dupire1997, DermanKani1994}, see \cite{Gatheral2006} for a review. We note that stochasticity of an additional variable entails additional parameters, particularly the centering and scale of the stochastic variable.

European Options do not heed the stochasticity of $r_2$, or, rather the differential between $r_1 - r_2$ as it is entirely inherited in the volatility of the forward $F = Se^{r_1 - r_2} t$, where $t$ is the time period, which can be captured by expert operators who typically use the volatility of the latter (at the nominal maturity) instead of that of the spot. However American options are specially --and seriously-- affected by both $r_1$ and $r_2$. In what follows, after discussing some empirical episodes, we will focus on injecting the dynamics of $r_1$ and $r_2$.
\begin{center}
**
\end{center}
The remaining part of this article is organized as follows. We present the real world problems as encountered by option operators in section \ref{episodes}, discuss their typology in section \ref{typology}, briefly link model error to fragility in section \ref{fragility}, present the master equation and the various possible dynamics and probability distributions for pricing in section in section \ref{sec:pricing}. We  perform a broad set of calculations under different models showing the robustness of our findings in the final sections. 

\section{Illustrative Practitioner Episodes}\label{episodes}
\textit{Practitioner Episode 1: The Currency Interest Rate Flip:}

Not only the sign of $(r_1-r_2)$ can vary, but it can flip from positive to negative --and the tradition for option pricing and hedging systems is to use a flag to price either the put or the call as if they were European since the early exercise feature can be ignored.

During the 1980s, German interest rates were generally lower than those in the United States. In such a configuration--where the foreign rate is below the domestic rate--standard pricing systems value the American put on a currency pair higher than its European counterpart, while assigning identical values to the corresponding calls. 

When interest rates later converged, and subsequently reversed following the post-reunification rise in German yields, many believed they were executing an arbitrage-free trade by selling the American option and buying the European one. Initially, their mark-to-model valuations appeared profitable, as the systems treated both options as equivalent. However, when interest rate differentials inverted, the mark-to-market values diverged dramatically. The models, which had ignored the early-exercise possibility of such options, failed to capture the exposure. Several trading desks incurred significant losses before realizing that the American call carried embedded optionality on the path of the rate differential.

Similar opportunities reappeared during subsequent currency crises and devaluations, whenever interest rates became unstable. The pattern was recurrent: volatility in the rate differential would amplify the hidden optionality of the American instrument, while the European remained constrained by its terminal payoff structure.

Option operators were unaware of the risks since both the academic literature and option software designers (an overlapping community) did not count for it --even stochastic volatility wasn't even implemented then prior to the late 1990s \cite{Taleb1997}.

\textit{Practitioner Episode 2: The Stock Squeeze:}
In the early 2000s, the corresponding author was confronted a problematic position: his desk was long listed American calls on an Argentinian stock and short the corresponding delta amount (hedge ratio) in the underlying shares. The stock, an obscure ADR, was delisted unexpectedly, forcing an urgent buy-in. No liquidity was available, and attempts to borrow the stock--ironically through the firm Bear Stearns at the time--proved futile.

The resolution was conceptually simple yet operationally decisive: exercise the calls up to the amount of the short (by the hedge ratio). By doing so, the trader obtained the shares and neutralized the squeeze. Had the options been European, early exercise would have been impossible, and the losses potentially catastrophic. The episode demonstrated that the American call possesses not only market optionality but also ``model error optionality''--the ability to adapt to unexpected discontinuities in the underlying or in the market microstructure. We note that such optionality can be modeled with a jump in the financing rate.

\textit{Practitioner Episode 3: The Equity Index Squeeze:}
A related mispricing witnessed by the corresponding author occurred in the period covering 1998-1999 (in the wake of the failure of the hedge fund Long Term Management). It concerned long-dated, over-the-counter European call options on an equity index. These instruments traded at prices corresponding to volatility levels far below any plausible historical measure. Traders were long the calls and short the index futures, continuously rebalancing as the market rose slowly but substantially. The problem is that the rebalancing led to an increase of short futures. They lost on the futures (which for these contracts were to be settled daily with an outflow of cash), but were unable to monetize gains on the options, which remained heavily discounted.

At one point, the options were offered below their intrinsic value relative to the forward (at a standard funding rate)--an apparent market inefficiency. Yet, capital constraints prevented arbitrage, as carrying the long position required margin and funding, not available to risky positions during that period. Earlier, during the crash of 1987, similar distortions were observed when the cash-futures discount widened to nearly 10\% --an arbitrage that failed to attract operators owing to the stress on the  financial system.

With European options, such dislocations can become terminal to a trading desk, that is, they threaten extinction. By contrast, the American contract provides a lower bound to adverse mark-to-market movements (and an option on funding rates): its early-exercise right effectively caps the degree of mispricing to which the holder can be exposed. This feature embodies an additional, often unrecognized, layer of convexity.
\section{Differential Valuation Cases}\label{typology}

\textit{Case 1: Convexity to Changes in the Carry:}
Consider an underlying forward and spot both initially at 100, and a one-year at-the-money European and American call. Under conventional pricing systems, both instruments will be marked identically.

If the underlying rallies to 140, both options converge to parity, each worth \$40. However, assume that interest rates rise to 10\%. The European option's value becomes the discounted intrinsic value--approximately \$36.36--while the American option, which can be exercised immediately, retains a value of \$40.

Thus, a change in the carry--here, the discounting environment--benefits the American option disproportionately. The European price is anchored to a fixed maturity, while the American's exercise flexibility preserves nominal value under higher rates.

\textit{Case 1B: Asymmetric Rate Shifts:}
Assume now that only the domestic rate increases to 10\%, with the spot unchanged at 140. The forward declines to roughly 126. The European call, valued off this forward, drops to approximately the present value of 26, or \$23.64. The American call, which may be exercised immediately, remains worth \$26.

In both scenarios, the American option systematically outperforms the European because it benefits from convexity to the interest rate differential. Any model that prices the two identically under changing carry assumptions is misspecified.

From this, a general principle follows: if option A is worth at least as much as option B in all scenarios, and strictly more in some, it is suboptimal to sell option A and buy option B at equal prices. Yet this qualitative inequality still leaves open the quantitative question--how much more should one pay for the flexibility?

\textit{Case 2: Sensitivity to Changes in the Foreign or Dividend Rate:}
Let $S = F = 140$ with both domestic and foreign rates initially at zero. Again, the European and American options start at the same model price. Suppose the foreign rate rises sharply to 20\%. The forward now appreciates to roughly $Se^{(r_d - r_f)T} \approx 1.16S$.

The European call, lacking early exercise, is now worth only about \$16 (its discounted intrinsic value). The American call, however, retains the full intrinsic value of \$40. The rationale is straightforward: the American option dynamically selects the more favorable exercise basis--cash or forward--depending on which maximizes its immediate payoff. It "chooses" the superior underlying, adapting endogenously to the change in rate environment.

\textit{Case 3: Sensitivity to the Yield Curve Slope:}
Consider now a non-flat term structure, such as those frequently observed around year-end or policy rollovers. When the yield curve contains inflection points, the conventional valuation using only the terminal forward rate $F_T$ becomes unreliable.

Intermediate fluctuations in the carry can significantly affect the American option's value, as the optimal exercise point may occur precisely at one of those kinks. A pricing or risk-management system that collapses the full term structure into a single terminal forward will therefore misprice the American option--often marking it equal to the European, when in fact it should be higher.

The intuition is clear: an American option allows the holder to "lock in" the forward at any intermediate date, capturing transient peaks in synthetic carry. The European option, constrained to final maturity, lacks such adaptability.

\section{Pricing Implementations}\label{sec:pricing}

The preceding examples illustrate that the value differential between American and European options grows with the volatility of interest rates and the curvature of the term structure. The greater the uncertainty in the path of the carry, the larger the unpriced optionality embedded in the American contract.

First, using the earlier notation in eq. $\ref{initial}$ we write down the price of an American option at time $0$ and underlying price $S$
\begin{equation}
O_A(S, K, \sigma, t, r_1, r_2),	= \sup_{\tau \in \mathcal{T}_{0,T}} \mathbb{E}^{\mathbb{Q}} \Bigl[ e^{-r_1 \tau} \, g(S_\tau) \Bigm] \label{mastereq}
\end{equation}
where $g(S)$  is the payoff function (intrinsic value) at exercise:
$g(S) = \max\bigl(\Phi (S - K), 0\bigr)$
where $\Phi = +1$ for a call option and $\Phi = -1$ for a put option,
$\mathcal{T}_{t,T}$  is the set of all stopping times $\tau$ such that $t \leq \tau \leq T$ almost surely,
$\mathbb{E}^{\mathbb{Q}}[\,\cdot]$ is the conditional expectation under the risk-neutral probability measure $\mathbb{Q}$, given the information available at time $t$ 
and, finally, $\mathbb{Q}$ is the risk-neutral (equivalent martingale) measure.
Now  eq. \ref{mastereq}, the "master" equation does not specify methodologies. 

Owing to the path dependence of American options, their pricing has always been fraught with difficulties, even in the very standard situation when only $S$ is stochastic. 

Note that conventional Monte Carlo methods are ill-suited to capturing this additional stochasticity, as the stopping time is path-dependent and endogenous. More sophisticated numerical approaches--such as least-squares Monte Carlo or hybrid analytical methods--are required to quantify the magnitude of this latent premium. In practice, however, even ordinal (directional) comparisons can reveal substantial model risk when early-exercise rights are ignored.

Some complexity arises from the uncertainty of the hedge horizon for the underlying. The effective forward hedge of an American option is unknown, since the exercise time is stochastic. The situation resembles that of a barrier option with an uncertain trigger: termination depends on multiple stochastic variables, including volatility, the base rate, and the rate differential.

A hidden risk arises from the following. Intuitively, the "smart" American option positions itself, in principle, at the point on the forward curve that maximizes its discounted value. A risk-management system that allocates all forward delta exposure to the terminal maturity--treating the forward as if exercise can occur only at $T$--commits a structural error. Such systems underestimate the embedded additional optionality and misstate both value and hedge sensitivities.

In summary, American options possess multiple layers of unaccounted convexity beyond their explicit early-exercise feature. These include sensitivity to stochastic rates, curvature in the term structure, model error, and liquidity constraints. Properly accounting for these requires stochasticizing the underlying rate processes and evaluating expected value under the distribution of exercise times--a problem intimately linked to the concept of the fugit.

\begin{remark}
	The difference between various methods should be minor compared to parameter uncertainty. We are looking for the first order effect of the stochasticity in rates, largely to gauge the magnitude of the hidden risk ignored so far.
	
	Risk management is about scenario analyses across a parameter set, not precise pricing; our approach allows parametrization.
	
	\end{remark}

\section{Integrating an American Option across Stochastic Rates}\label{sec:integration}
In short, in what follows, we try the following simplified heuristics to grasp the hidden exposure. All are based on a separation of $O_A$ using a separation of the sort used in eq. \ref{eqforfragility}, that is integrating $O_A$ across $r_1$ or $r_2$.
\begin{mdframed}
	\begin{itemize}
	\item Method 1- One single integration $O_A$ across stochastic rates at a distribution of optimal stopping times, the " fugit based heuristic".
	\item Method 2- Multiple integrations of $O_A$ across stochastic rates at a given optimal stopping time $\tau$, the "fugit".
\end{itemize}
\end{mdframed}

Let us use the shortcut $O_A(r_1, r_2, t)$ to denote the price of an American option computed under a deterministic carry and foreign or dividend rates $r_2$ using any standard numerical method (binomial, lattice, or PDE). We wish to approximate the price of the same option when either $r_1$ or $r_2$ is stochastic, by integrating over the distribution of the stochastic rate(s) at the \textit{effective exercise time}.
We note that perturbations for $r_1$ can cover squeezes of financing (in section \ref{typology}), while $r_2$ can cover changes in the security yield, which includes dividends.

\subsection{Various stopping times methodologies}
We first proceed by assuming that one of the rates is stochastic, then expand for both assuming either independence or some correlation between the rates.
\subsubsection{Single expected Stopping Time}\label{sec:heuristics}

The  "Fugit"-based Heuristic, see \cite{Taleb1997}, is as follows. Let $\tau^*$ be the expected discounted stopping time of the American option, measured in risk-neutral time units. If $\tau$ denotes the random stopping time (optimal exercise), then the fugit is defined as

$$
\tau^*(S_0, t_0) = \mathbb{E}^Q \left[ \int_{t_0}^{\tau} e^{-r(u - t_0)} \, du \right],
$$
which can be interpreted as the "effective maturity" or the time-to-exercise that discounts equivalently to the American payoff. For European options, $\tau^* = T - t_0$; for deep-in-the-money Americans, $\tau^*$ is substantially shorter.

This quantity can be estimated directly from a binomial or finite-difference grid as the expectation of discounted time spent before exercise.

A trick is proposed by \cite{Taleb1997} as a "shortcut method... to find the right duration (i.e., expected time to termination) for an American option". Taleb terms this result "Omega". The formula is

\begin{equation}
\Omega = t \frac{\rho_{2A}}{\rho_{2E}},
\label{eq:heuristic}
\end{equation}
where $t$ is the nominal time to expiration, $\rho_{2A}$ and $\rho_{2E}$ are "Rhos", the sensitivities of the American and European options to changes in the underlying nominal carry yield.

\subsubsection{The Stochastic Fugit: Distribution of Exercise Times}\label{sec:fugit}
A deterministic-rate American pricer (binomial, finite-difference, or least-squares Monte Carlo) naturally yields:
\begin{itemize}
    \item a discrete set of candidate exercise times $t_1, \ldots, t_K$,
    \item the corresponding exercise probabilities $p_k = \mathbb{P}(\tau = t_k)$.
\end{itemize}
This defines the \emph{stochastic fugit}:
\[
\tilde{T} \in \{t_1, \ldots, t_K\},
\qquad
\mathbb{P}(\tilde{T} = t_k) = p_k.
\]
The classical deterministic fugit is merely the expectation
\begin{equation}
    \tau^* = \mathbb{E}[\tilde{T}] = \sum_{k=1}^{K} p_k t_k.
    \label{eq:expectedvalue}
\end{equation}

By retaining the full distribution $(t_k, p_k)$, we preserve the time convexity inherent in the early-exercise feature.

The fugit provides a principled estimate of the \textit{expected exercise horizon} for use in the rate distribution. It adjusts automatically to the option's moneyness. This heuristic captures first-order effects of rate uncertainty without solving a full two-factor PDE. It can be extended by integrating over a discrete distribution of fugit times $t_k$ from a Bermudan exercise histogram,
\begin{equation}
O_A \approx \sum_k p_k \int O_A(r, t) \, f_{r(t_k)}(r) \, dr,
\label{eq:fulldistribution}
\end{equation}

where $p_k = \mathbb{P}(\tau^* = t_k)$.
\subsubsection{Fugit-weighted integration heuristic}\label{sec:fugitintegration}

We define the fugit-weighted American price under rate stochasticity as

\begin{equation}
O_{A, \tau^*} = \int_{\mathbb{D}_{r}} O_A(r, t) \, f_{r(\tau^*)}(r) \, dr,
\label{eq:integration}
\end{equation}
where $f_{r(\tau^*)}$ is the density of the stochastic rate evaluated at the expected stopping time $\tau^*$. This represents a weighted average of deterministic-rate American prices, with the weights given by the probability distribution of the relevant rate at the fugit time.

\subsection{Extension to Two Stochastic Rates}\label{sec:tworates}
When both rates $r_1(t)$ and $r_2(t)$ are stochastic, possibly correlated, the extension is immediate:
\begin{multline}
O_A^{(\text{stoch-fugit})}=\\
\sum_{l=1}^{L} p_l
\iint_{\mathbb{R}^2}
O_A(r_1, r_2, t_l)\,
f_{(r_1,r_2)(t_l)}(r_1, r_2)\,
dr_1\,dr_2	.
\end{multline}

If independence is assumed,
\[
f_{(r_1,r_2)(t_k)}(r_1, r_2)
=
f_{r_1(t_k)}(r_1)\,
f_{r_2(t_k)}(r_2).
\]
\bigskip

\subsection{Various distribution of rates}\label{sec:ratedistributions}

Let the funding rate $r_1(t)$ or carry rate $r_2(t)$ follow one of the canonical short-rate dynamics:

\textit{a) Bachelier or normal world.:}
\[
dr = \mu_r \, dt + \sigma_r \, dW_t
\]
\[
\Rightarrow \quad r_{\tau^*} \sim \mathcal{N}(r_0 + \mu_r \tau^*, \, \sigma_r^2 \tau^*).
\]
\textit{b) Vasicek / Hull--White world.:}
\[
dr = \kappa_r (\theta_r - r) \, dt + \sigma_r \, dW_t
\]
\[
\Rightarrow
\begin{cases}
\mathbb{E}[r_{\tau^*}] = \theta_r + (r_0 - \theta_r) e^{-\kappa_r \tau^*}, \\
\mathrm{Var}[r_{\tau^*}] = \dfrac{\sigma_r^2}{2\kappa_r} (1 - e^{-2\kappa_r \tau^*}).
\end{cases}
\]

\textit{c) Lognormal world.:}
$$
\frac{dr_2}{r} = \mu_r \, dt + \sigma_r \, dW_t
$$
\begin{multline*}
	\Rightarrow \quad r_{\tau^*} = r_{20} \exp\!\left( (\mu_r - \tfrac{1}{2}\sigma_r^2)\tau^* + \sigma_r \sqrt{\tau^*} \, Z \right),\\
\qquad Z \sim \mathcal{N}(0,1).
\end{multline*}

\bigskip
\begin{center}
**
\end{center}

The end result for us is testing , where $\widetilde{.}(r)$ denotes stochasticity over parameter $r$, $\widetilde{O_A(r)} -O_A(r)$, the extra optionality, after clearing a few hurdles.

We will perform tests to establish whether the fugit shortcut represents a good enough an approximation and whether various rate dynamics (presented in the next section) make a difference for the convexity bias.

\begin{figure} [h!]
\includegraphics[scale=0.5]{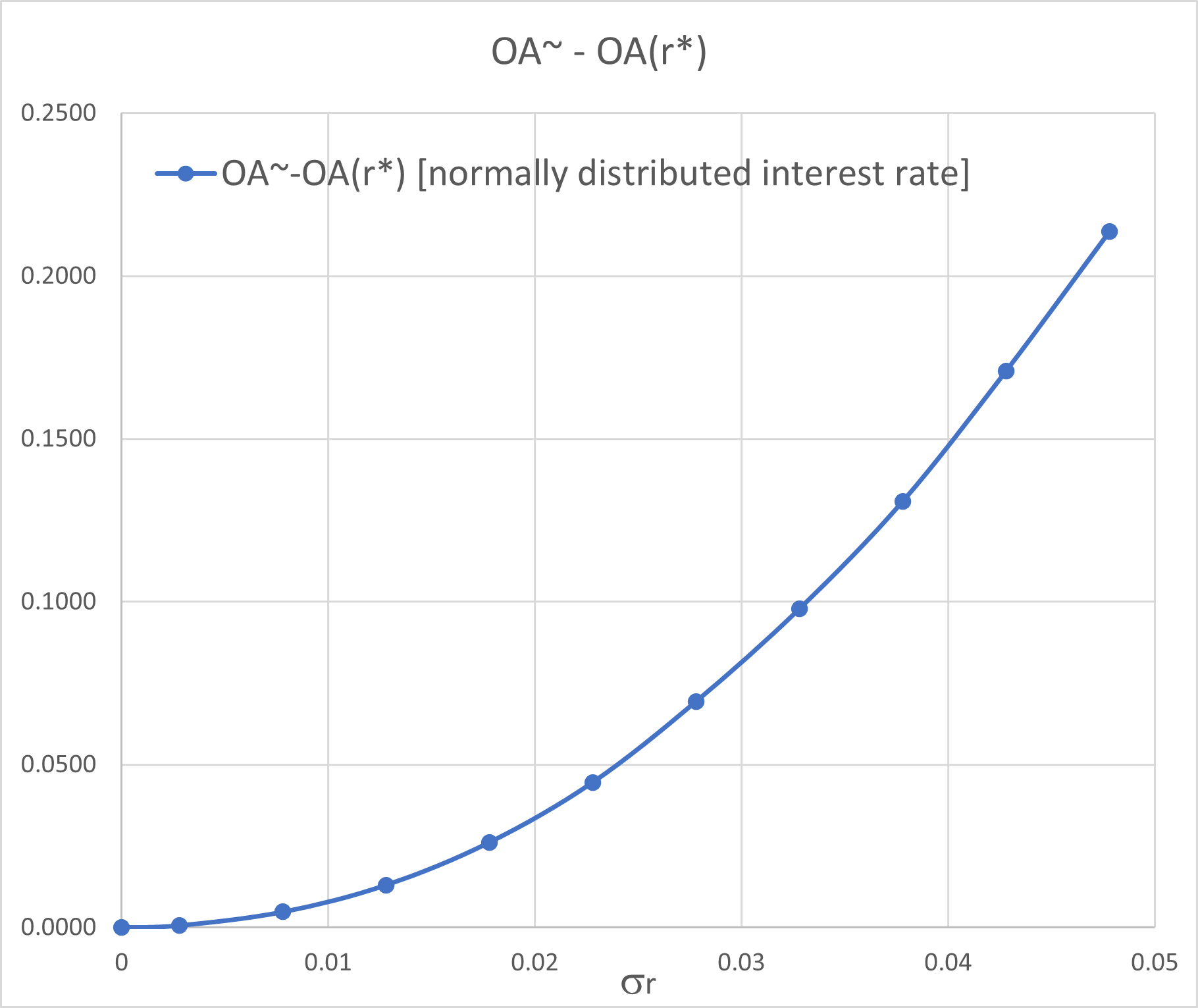}
  \caption{Optionality versus standard deviation for a an equity put option under a normally distributed local interest rate with $OA(r^*)=14.3184$, $S/K=1.00$ }\label{fig:sigma1}
  \end{figure}
  
\begin{figure} [h!] 
 \includegraphics[scale=0.5]{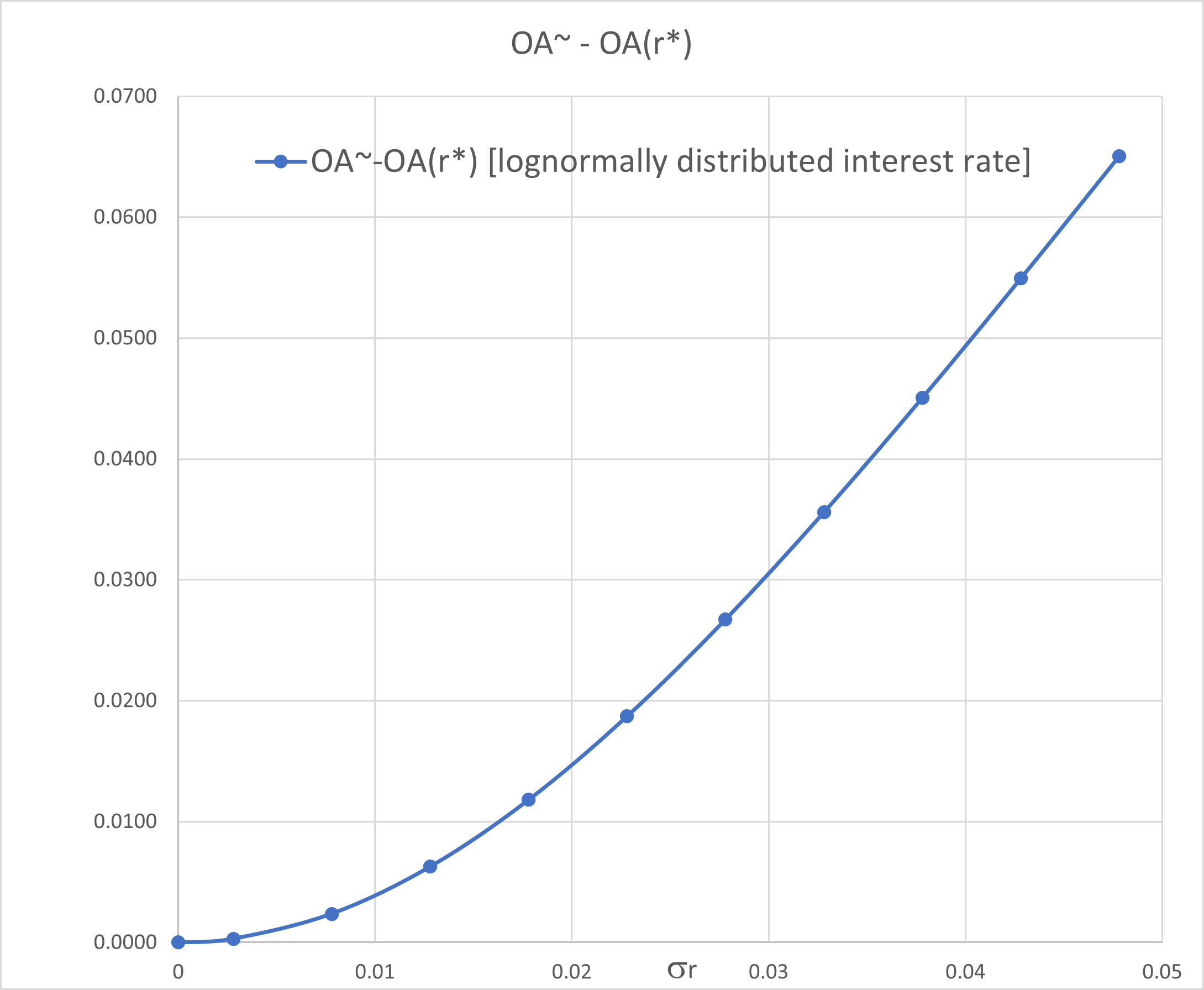}
  \caption{Optionality versus standard deviation for a currency put option under a lognormally distributed local interest rate with $OA(r^*)=15.1700$ and $S/K=1.00$}\label{fig:sigma2}
  \end{figure}
  
\begin{figure} [h!]
 \includegraphics[scale=0.5]{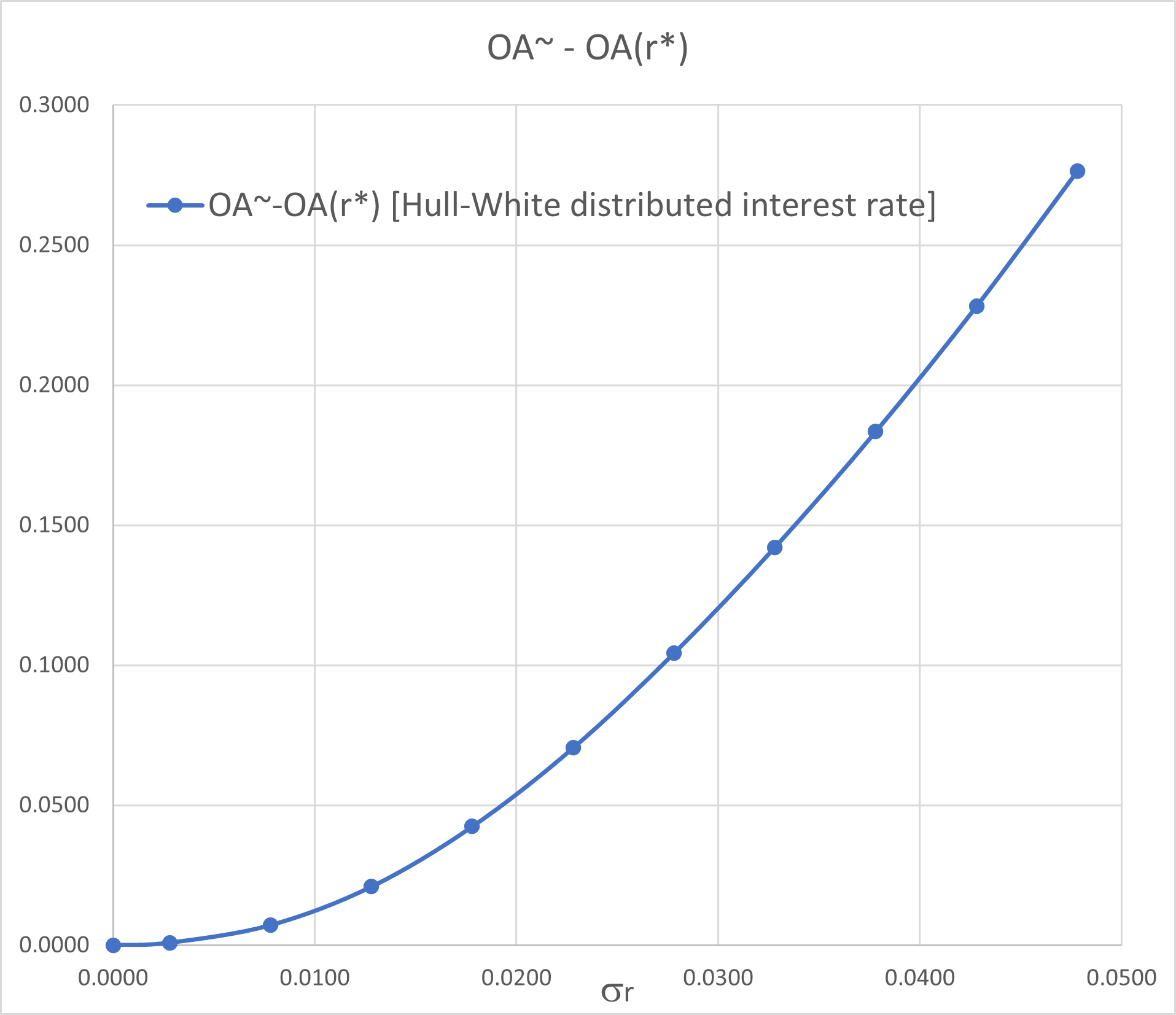}
  \caption{Optionality versus standard deviation for a currency call option under a Hull-White distributed local interest rate with $OA(r^*)=12.7779$ and $S/K=1.00$}\label{fig:sigma3}
  \end{figure}

\section{Main Numerical Implementation}\label{sec:numexperiments}
We work throughout these simulations with options with --to normalize --a maturity of one year, hence no loss of generality. We consider equity puts, currency puts, and currency calls with stochastic local rate.

In our base example, we consider a generic at-the-money American equity put option on a high volatility stock, with the following common parameters, volatility (put on the upper end of common values), $\sigma=40\%$, maturity, $T=12$ months, initial underlying asset value, $S_0=100$, strike price, $K=100$, and a stochastic interest rate with an initial value $r_0=1\%$. (With respect to the notation in Section I of this paper, this is an American option with $r_1=r$ and $r_2=0$.)  For the stochastic  rate, we assume that it follows a Bachelier process (that is, normally distributed as defined in Section \ref{sec:ratedistributions}), with mean $\bar{r}=4.18\%$ and standard deviation $\sigma_r=1.28\%$ at maturity, as described in Section \ref{sec:ratedistributions} of this paper.  These parameters align with recent values of the 1-year US treasury rate. Specifically, we consider the 1-year treasury yield over the past three years \cite{FRED2025}, where the yield is reported at the end of every month from January 2022 until March 2025. 


We start by computing the stopping time using expectation of the classical deterministic fugit using \eqref{eq:expectedvalue}. Accordingly, we compute the American option value, $\widetilde{O}_A(r)$,  under a normal interest rate using \eqref{eq:integration}.  The parameters of the interest rate process are determined based on matching the first two moments of the rate at maturity, $r_T$, with $\bar{r}$ and $\sigma_r$. For instance, under a normally distributed interest rate, the drift and volatility are  obtained by moment matching as $\mu=\frac{\bar r - r_0}{T}$ and $\sigma = \sigma_r/\sqrt{T}$. The binomial lattice is used to compute the American option price for a fixed interest rate, taking into consideration the early exercise feature. Then, using the Gauss-Hermite quadrature  we approximates the expectation of this price with respect to the stochastic interest rate by evaluating the lattice-based price at a finite number of carefully chosen interest rate realizations and aggregating them using predetermined weights. More details about the approach can be found in section \ref{further}.

We compare the results with the corresponding deterministic price of an American option, with the local rate being equal to the average interest rate at the expected fugit $\tau^*$, $O_A(r^*)$. We then estimate 
\begin{equation}\label{eq:optionality}
\pi_A=\widetilde{O}_A(r)-O_A( r^*), 
\end{equation}
as a measure of the gain from the hidden optionality of the American option. We compute $\pi_A$ and present the results in Table \ref{tab:sigma1}. To investigate the impact of interest-rate uncertainty, we vary the standard deviation parameter of the interest-rate process $\sigma_r$.  As expected, the difference $\pi_A$ increases with the variability of the interest rate. When the volatility is set to zero, the stochastic model converges to the deterministic case and the difference becomes exactly zero. This monotonic increase in $\pi_A$ as standard deviation rises is clearly observed in Figure \ref{fig:sigma1}. We repeat the same numerical experiments under a local rate following Geometric Brownian motion and Hull-White processes, as defined in Section \ref{sec:ratedistributions}, and present the results in Figures \ref{fig:sigma2} and \ref{fig:sigma3}, respectively. We observe a similar behavior, as discussed in more details in section \ref{further}.
\begin{table}[h!]
\centering
\small
\renewcommand{\arraystretch}{0.6} 
\begin{tabular}{|c|c|c|}
\hline
$\sigma_r$ & \textbf{$\widetilde{O}_A(r)$}& \textbf{$\pi_A$}  \\
\hline
 0.0000 & 14.3184 & 0.0000 \\ \hline
 0.0028 & 14.3190  & 0.0006 \\ \hline
0.0078 & 14.3231  & 0.0048 \\ \hline
 0.0128 & 14.3314  & 0.0130 \\ \hline
0.0178 & 14.3445 & 0.0261 \\ \hline
0.0228 & 14.3629  & 0.0446 \\ \hline
0.0278 & 14.3878  & 0.0694 \\ \hline
 0.0328 & 14.4162  & 0.0978 \\ \hline
0.0378 & 14.4492  & 0.1308 \\ \hline
0.0428 & 14.4892  & 0.1708 \\ \hline
 0.0478 & 14.5319& 0.2136 \\ \hline
\end{tabular}
\caption{Optionality as a function of the standard deviation of a normally distributed interest rate with $OA(r^*)=14.3184$, $S/K=1.00$}\label{tab:sigma1}
\end{table}

\section{Further Numerical Results and Technical Details}
In Section \ref{BI}, we briefly discuss the binomial lattice method (\cite{Cox}) used to compute the integration and to obtain the expected fugit. In Section \ref{BII}, we investigate the proposed heuristics in \eqref{eq:heuristic} and then compare the results with the expected fugit estimated from the lattice in Section \ref{BI}, which can be seen as the ``exact" baseline.  In Section \ref{BIII}, we study how changes in the dynamics of interest rate, under alternative distributional assumptions, affect our results. Finally, In Section \ref{BIV}, we briefly exhibit an alternative approach of measure optionality by comparing the American option value to a European counterpart.  

\subsection{Binomial Lattice and Fugit Distribution}\label{BI}
This section presents the numerical framework used throughout the paper. The starting point is the computation of $\tau^* = \mathbb{E}[\tilde{T}]$ in \eqref{eq:expectedvalue}, the expected optimal stopping time of an American option using the classical deterministic fugit. Then, we employ the binomial lattice to compute the value of an American option with stochastic local rate, $\widetilde{O}A(r)$ , through numerical integration. 

Following the notation described in \cite{ElHassanandMaddah}, we briefly recall the elements of the lattice that are required for the computation of the expected fugit and the subsequent numerical valuation. Let the maturity $T$ be divided into $n$ time steps of length $\delta t = T/n_s$. In accordance with \cite{ElHassanandMaddah}, we set $n=2000$, which provides stable and accurate numerical results. The stock price evolves on the lattice according to the multipliers $u = e^{\sigma \sqrt{\delta t}}$ and $d = \frac{1}{u}$, with a risk--neutral probability $q = \frac{e^{(r_f-\delta)\delta t} - d}{u - d}$, where $r_f$ denotes the continuously compounded risk--free rate and $\delta$ the dividend yield (or foreign rate in the currency case). Starting with $S(i,j)$, the stock price at time step $i$ and state $j$, and by $P(i,j)$ the American option value can be obtained through backward recursion. At maturity, $i = n$, the final payoff is $P(n_s,j) = \max(K - S(n_s,j),\,0)$ for a put option (with an analogous expression for a call). For $i < n$, the value of the option is the maximum between the exercise payoff and the continuation value, $P(i,j) = \max\!\bigl\{ K - S(i,j),\; V_c(i,j) \bigr\}$, where the continuation value is given by $V_c(i,j)=e^{-r_f\delta t}\Bigl[ q\,P(i+1,j+1) + (1-q)\,P(i+1,j) \Bigr]$. The backward recursion simultaneously determines the optimal exercise region. Accordingly, we can define an early exercise indicator
\[
I(i,j) =
\begin{cases}
1, & \text{if } K - S(i,j) \ge V_c(i,j), \\
0, & \text{otherwise}.
\end{cases}
\]
The optimal stopping time is therefore the first index
\[
\tau = \min\{ i \ge 0 : I(i,j_i) = 1 \},
\]
with corresponding point in time $t^* = \tau\,\delta t$.

To compute the distribution of $\tau$, the probability mass function is propagated forward through the lattice while enforcing the optimal stopping rule. Let $\pi(i,j)$ denote the probability of reaching node $(i,j)$ without
prior exercise, with initialization $\pi(0,0)=1$. For each time step $i=0,\dots,n_s-1$, the below strategy is followed
\begin{itemize}
\item If $I(i,j)=1$, the probability $\pi(i,j)$ is recorded as stopping at time index $i$ and is not propagated further.
\item If $I(i,j)=0$, the probability evolves according to the risk--neutral dynamics,
\begin{multline}
\pi(i+1,j+1) =\pi(i+1,j+1) + q\,\pi(i,j), \\
\pi(i+1,j) =\pi(i+1,j)+ (1-q)\,\pi(i,j).
\end{multline}

\end{itemize}

Denoting by $P_f(i)$ the probability of optimal exercise at step $i$, the expected fugit conditional on exercise is
\[
\tau^*=\mathbb{E}[t^* \mid \text{exercise}]
=
\frac{\sum_{i=0}^{n_s} (i\,\delta t)\,P_f(i)}
{\sum_{i=0}^{n_s} P_f(i)}.
\]

A complete algorithm to compute the fugit pmf $P_f(i)$ and expected value $\tau^*$  is presented next.   While the algorithm is straightforward, we could not identify any similar approaches in the  literature.  There are some online forums hinting to it without enough details.  The following complete algorithm can be seen as a side-contribution of this paper. 
\medskip

\noindent
\textbf{Algorithm for the fugit distribution.}

{\small

\noindent\textbf{\texttt{Step 1}}\\
\texttt{Input the values for $S_0$, $K$, $T$, $n$, $R$, $\delta$, and $\sigma$.}\\
\texttt{Set $\delta t = T/n_s$. Construct the CRR parameters $u$, $d$, and $q$ (see \cite{ElHassanandMaddah}.}\\

\noindent\textbf{\texttt{Step 2}}\\
\texttt{Build the recombining stock-price lattice $\{S(i,j)\}_{0\le i\le n,\,0\le j\le i}$.}\\

\noindent\textbf{\texttt{Step 3}}\\
\texttt{Initialize the terminal option values at maturity:}\\
\hspace{0.6cm}\texttt{For $j=0,\dots,n$, set $P(n,j)=\Phi(S(n,j))$, where $\Phi(S)=K-S$.}\\

\noindent\textbf{\texttt{Step 4}}\\
\texttt{Backward recursion (pricing + exercise indicator):}\\
\hspace{0.6cm}\texttt{For $i=n-1,\dots,0$:}\\
\hspace{1.2cm}\texttt{For $j=0,\dots,i$:}\\
\hspace{1.8cm}\texttt{Set $V_c(i,j)=\frac{1}{R}\Big(q\,P(i+1,j+1)+(1-q)\,P(i+1,j)\Big)$.}\\
\hspace{1.8cm}\texttt{Set $P(i,j)=\max\{\Phi(S(i,j)),\,V_c(i,j)\}$.}\\
\hspace{1.8cm}\texttt{If $\Phi(S(i,j))\ge V_c(i,j)$ set $I(i,j)=1$, else set $I(i,j)=0$.}\\

\noindent\textbf{\texttt{Step 5}}\\
\texttt{Forward recursion (probabilities of reachable nodes and stopping distribution):}\\
\hspace{0.6cm}\texttt{Create an empty matrix $\pi$ of size $(n+1)\times(n_s+1)$ and set $\pi(0,0)=1$.}\\
\hspace{0.6cm}\texttt{Create an empty vector $P_f$ of length $(n_s+1)$ and set $P_f(i)=0$ for all $i$.}\\
\hspace{0.6cm}\texttt{For $i=0,\dots,n-1$:}\\
\hspace{1.2cm}\texttt{For $j=0,\dots,i$:}\\
\hspace{1.8cm}\texttt{If $\pi(i,j)=0$, continue.}\\
\hspace{1.8cm}\texttt{If $I(i,j)=1$, set $P_f(i)=P_f(i)+\pi(i,j)$ (stop here).}\\
\hspace{1.8cm}\texttt{Otherwise propagate:}\\
\hspace{2.4cm}\texttt{$\pi(i+1,j+1)=\pi(i+1,j+1)+q\,\pi(i,j)$,}\\
\hspace{2.4cm}\texttt{$\pi(i+1,j)=\pi(i+1,j)+(1-q)\,\pi(i,j)$.}\\

\noindent\textbf{\texttt{Step 6}}\\
\texttt{Maturity handling:}\\
\hspace{0.6cm}\texttt{For $j=0,\dots,n$, if $\pi(n,j)>0$ and $\Phi(S(n,j))>0$,}\\
\hspace{1.2cm}\texttt{set $P_f(n)=P_f(n)+\pi(n,j)$.}\\

\noindent\textbf{\texttt{Step 7}}\\
\texttt{Expected fugit (conditional on exercise):}\\
\hspace{0.6cm}\texttt{Set $t_i=i\,\delta t$ for $i=0,\dots,n$.}\\
\hspace{0.6cm}\texttt{Set $E_t=\dfrac{\sum_{i=0}^{n} t_i\,P_f(i)}{\sum_{i=0}^{n} P_f(i)}$.}\\

\noindent\textbf{\texttt{Output}}\\
\texttt{Return the stopping distribution $\{P_f(i)\}_{i=0}^{n}$ and the expected fugit $E_t$.}

}

Once the expected fugit has been obtained, the same binomial lattice is used to evaluate the optimality measure $\pi_A$. First, a deterministic benchmark is computed by evaluating the lattice with an interest rate $r^* = {E}[r(t^*)]$, . For instance, under a normally distributed interest rate, the deterministic benchmark rate is simply ${E}[r(t^*)]=r_0+\mu \tau^*$. 
Having defined the deterministic value $OA(r^*)$, the optimality measure $\pi_A$ is obtained by comparing $OA(r^*)$ with the option value that accounts for a stochastic interest rate, $\widetilde{O}A(r)$. Accordingly, $\widetilde{O}A(r)$ is obtained using the obtained expected fugit and the integration in \eqref{eq:integration}, which is evaluated numerically using the Gauss-Hermite quadrature (e.g.~\cite{StoerBulirsch2013}). This numerical integration method approximates the expectation by a weighted sum of lattice prices evaluated at optimally chosen rate nodes.  

\subsection{Expected Sopping Time $\Omega$}\label{BII}
In this section, we compare the effective stopping time developed by Taleb \cite{Taleb1997} using \eqref{eq:heuristic} and using the expectation of the classical deterministic fugit in \eqref{eq:expectedvalue}, for an equity put, currency put, and currency call. 
\subsubsection*{B.II.1 Equity Put}
\begin{figure}[h!]
\includegraphics[scale=0.5]{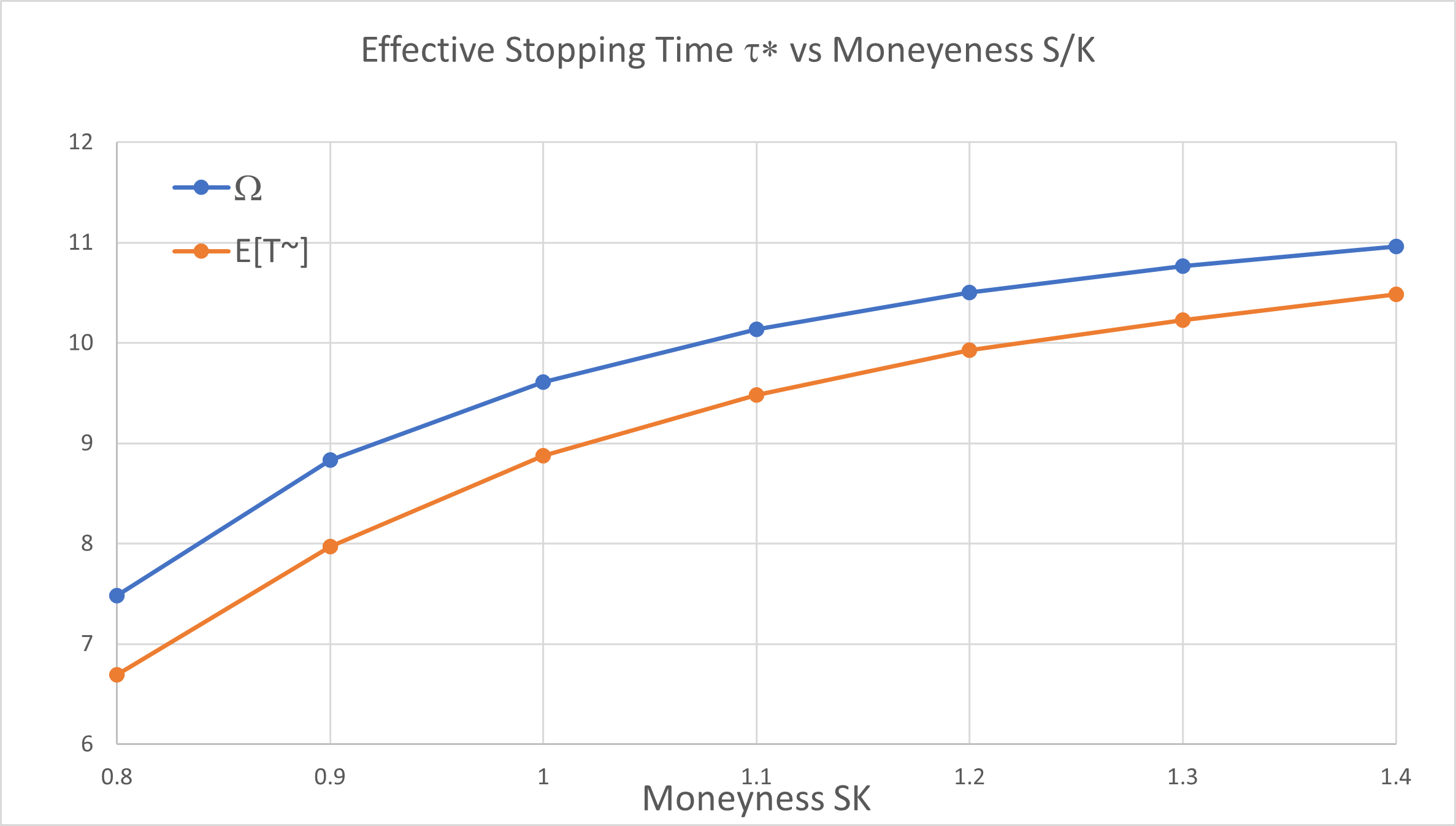}
  \caption{Effective sopping time $\tau^*$ versus moneyness for an equity put option.}\label{fig:exercisetime}
  \end{figure}

We consider an equity put option with the input parameters mentioned in Table \ref{tab:sigma1}. 
We estimate $O_A(r)$ from the binomial lattice method, at the deterministic benchmark, as explained in Section B.I. The European call $O_E(r)$ is valuated based on the Black-Scholes-Merton formula (\cite{Black}).  We first estimate the American and European option sensitivities with respect to the interest rate, $\rho_{2A}=\frac{\partial O_A(r)}{\partial r}$ and $\rho_{2E}=\frac{\partial O_E(r)}{\partial r}$, using a common heuristic based on central difference method, 
\begin{multline*}
\rho_{2J}=\\
\frac{|O_A(r+0.01,t)-O_A(r,t)|+|O_A(r-0.01,t)-O_A(r,t)|}{2},\\
 J=A,E.
\end{multline*}
We use the values of $\rho_{2A}$ and $\rho_{2E}$ to obtain the effective stopping time associated with the early exercise feature of an American option using \eqref{eq:heuristic}. We also compute compute the stopping time using expectation of the classical deterministic fugit using \eqref{eq:expectedvalue}. We obtain   close results, as shown in Table \ref{tab:exercisetime}, which confirm the validity of the heuristics approached suggested by Taleb \cite{Taleb1997}. The results also show how $\tau^*$ changes with moneyness. As expected, a lower moneyness level implies an earlier exercise time, i.e., the expected fugit is increasing in the moneyness.   Figure \ref{fig:exercisetime} also demonstrates this graphically.
\begin{table}[h!]
\centering
\small
\renewcommand{\arraystretch}{0.6} 
\begin{tabular}{|c|c|c|c|c|}
\hline
\textbf{$S/K$} & \textbf{$\rho_A$} & \textbf{$\rho_E$} & \textbf{$\tau^*=\Omega$} & \textbf{$\tau^*=\mathbb{E}[\tilde{T}]$} \\
\hline
$0.8000$ & $0.29931$ &$0.4800$&$ 7.4833$&$ 6.6954$ \\
\hline
$0.9000$ & $0.3204$ & $0.4353$&$8.8335$ &$7.7924$ \\
\hline
$1.0000$ & $0.3047$ & $0.3804$ &$ 9.6117$&$8.8781$ \\
\hline
$1.1000$ & $0.2728$ & $0.3230$&$ 10.1349$ &$9.4785$ \\
\hline
$1.2000$& $0.2348$ & $0.2682$ &$ 10.5033$&$9.9259$ \\
\hline
$1.3000$& $0.1964$ & $0.2190$ &$ 10.7641$&$10.2288$ \\
\hline
$1.4000$& $0.1611$ & $0.1764$ &$ 10.9622$ & $10.4821$\\
\hline
\end{tabular}
\caption{Effective stopping time $\tau^*$, in months, for an equity put option.}\label{tab:exercisetime}
\end{table}

\
\subsubsection{Currency Put}\label{BII2} 
In this section, we consider a currency put with base parameter values similar to those in section \ref{}, with a foreign rate $r_2=2.8\%$. These parameters are given in Table \ref{tab:inputparameterscurput} for completeness.

\begin{table}[h!]\label{tab:stocrbase}
\centering
\small
\renewcommand{\arraystretch}{0.6} 
\begin{tabular}{|c|c|c|c|c|c|c|c|}
\hline
$K$ & $S$ & $T$ & $\sigma$ &$r_2$&$r_{10}$ &$\bar{r_1}$ & $\sigma_{r_1}$ \\
\hline
100 & 80 & 1y & 40\% & 2.8000\%&1\%&4.18\% & 1.28\% \\
\hline
\end{tabular}
\caption{Input parameters for the base example on interest stochasticity for a currency put} \label{tab:inputparameterscurput}
\end{table}

\begin{figure} [h!]
 \centering
   
 \includegraphics[scale=0.4]{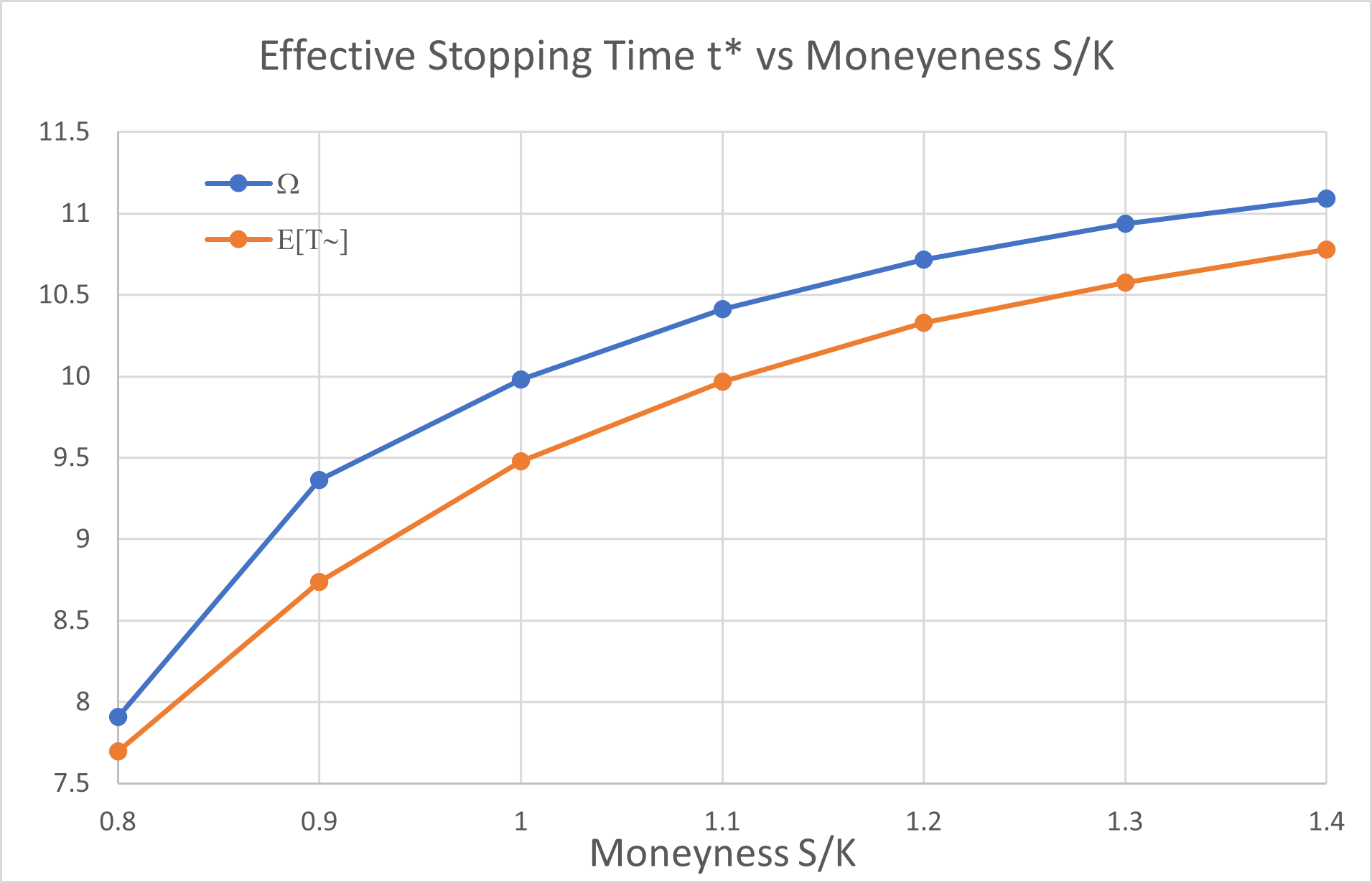}
  \caption{Effective sopping time $\tau^*$ versus moneyness for an American currency put }\label{fig:exercisetimecurrencyput}
  \end{figure}

Similar to equity puts, we determine the effective stopping time $\tau^*$ for the American currency put using \eqref{eq:heuristic} and \eqref{eq:expectedvalue}. The results are presented in 
Figure \ref{fig:exercisetimecurrencyput} for different moneyness levels. We obtain again similar results as in the case of an equity put option confirming the validity of Taleb's heuristic in \eqref{eq:heuristic}.  

\begin{table}[h!]
\centering
\small
\renewcommand{\arraystretch}{0.6} 
\begin{tabular}{|c|c|c|c|c|}
\hline
\textbf{$S/K$} & \textbf{$\rho_A$} & \textbf{$\rho_E$} & \textbf{$\tau^*=\Omega$} & \textbf{$\tau^*=\mathbb{E}[\tilde{T}]$} \\
\hline
$0.8000$ & $0.4744$ &$0.4876$&$8.4816$&7.6977 \\
\hline
$0.9000$ & $0.4380$ & $0.4477$&$9.4292$&8.7379  \\
\hline
$1.0000$ & $0.3904$ & $0.3960$ &$ 10.0379$&9.4770 \\
\hline
$1.1000$ & $0.3371$ & $0.3404$&$ 10.4487$&9.9653  \\
\hline
$1.2000$& $0.2841$ & $0.2859$ &$ 10.7387$&10.3291 \\
\hline
$1.3000$& $0.2349$ & $0.2359$ &$ 10.9521$&10.5744 \\
\hline
$1.4000$& $0.1915$ & $0.19215$ &$ 11.1036$&10.7799 \\
\hline
\end{tabular}

\caption{Effective sopping time $\tau^*$ versus moneyness for a currency put option.}\label{tab:exercisetimecurrencyput}
\end{table}

\vspace{-1 mm}

\subsubsection{Currency Call}\label{BII3}
\begin{figure} [h!]
 \centering
   
 \includegraphics[scale=0.4]{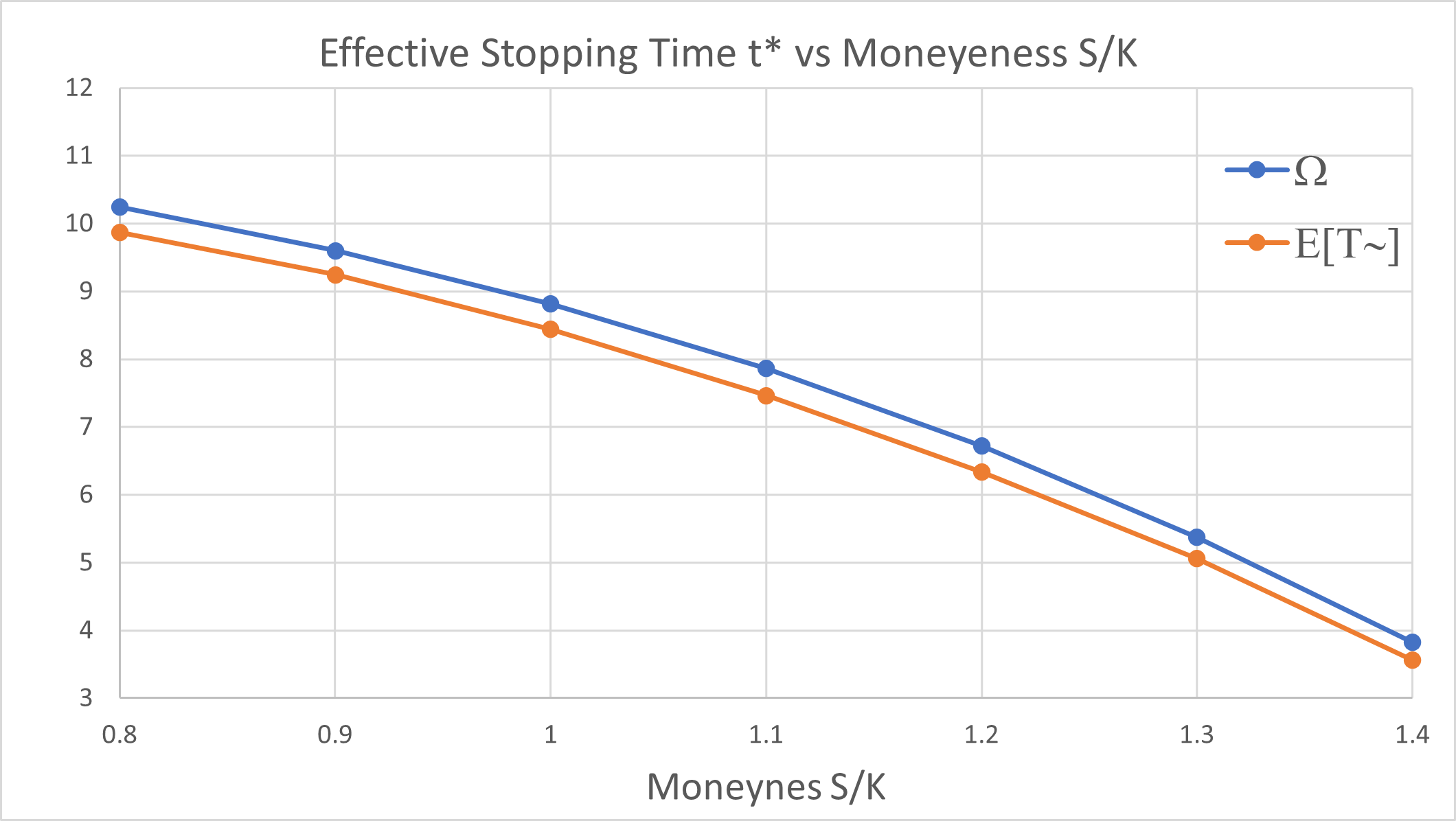}
  \caption{Effective sopping time $\tau^*$ versus moneyness for an American currency call }\label{fig:currencycall}
  \end{figure}

We study a currency call option with parameters similar to those of the currency put above but with a foreign rate $r_2=10\%$. The observations we make here are also applicable to call options with dividends, which have a similar pricing structure.\footnote{Optionality in the context of non-dividend paying equity calls is not relevant as it is not optimal to exercise these options before maturity, e.g.~Hull \cite{{HullBook}}.} The results in 
Figure \ref{fig:currencycall}, which again  validates the heuristic \eqref{eq:heuristic}.  

\begin{table}[h!]
\centering
\small
\renewcommand{\arraystretch}{0.6} 
\begin{tabular}{|c|c|c|c|c|c|c|c|}
\hline
$K$ & $S$ & $T$ & $\sigma$ &$r_f$&$r_0$ &$\bar{r}$ & $\sigma_r$ \\
\hline
100 & 80 & 1y & 40\% & 10\%&3\%&4.18\% & 1.28\% \\
\hline
\end{tabular}
\caption{Input parameters for the base case for currency call option under stochastic local rate}\label{tab:inputparameters1}
\end{table}

\begin{table}[h!]
\centering
\small
\renewcommand{\arraystretch}{0.6} 
\begin{tabular}{|c|c|c|c|c|}
\hline
\textbf{$S/K$} & \textbf{$\rho_A$} & \textbf{$\rho_E$} & \textbf{$\tau^*=\Omega$} & \textbf{$\tau^*=\mathbb{E}[\tilde{T}]$} \\
\hline
$0.8000$ & $0.1900$ &$0.2226$&$ 10.24196$&9.8729 \\
\hline
$0.9000$ & $0.2719$ & $0.3399$&$9.60074$&9.2454  \\
\hline
$1.0000$ & $0.3470$ & $0.4722$ &$ 8.8180$&8.4378 \\
\hline
$1.1000$ & $0.4011$ & $0.6123$&$ 7.8608$&7.4642  \\
\hline
$1.2000$& $0.4226$ & $0.7548$ &$ 6.7186$ &6.3336\\
\hline
$1.3000$ & $0.4013$ & $0.8955$&$ 5.3769$ &5.0576 \\
\hline
$1.4000$& $0.3290$ & $1.0322$ &$3.8250 $ &3.5660\\
\hline
\end{tabular}
\caption{Effective stopping time $\tau^*$ for an American currency call option}\label{tab:currencycall}
\end{table}

\subsection{ Different Rate Dynamics and Moneyness Levels}\label{BIII}
In this section, we evaluate the hidden optionality of American equity puts, currency puts, and currency calls, under different rate dynamics. Similar to section \ref{sec:numexperiments}, we start by considering a normally distributed interest rate, and then extend the analysis to a lognormally and Hull-White distributed rate.

\subsubsection{Equity Put}\label{BIII1} 
\begin{figure} [h!]
 \centering
   
 \includegraphics[scale=0.5]{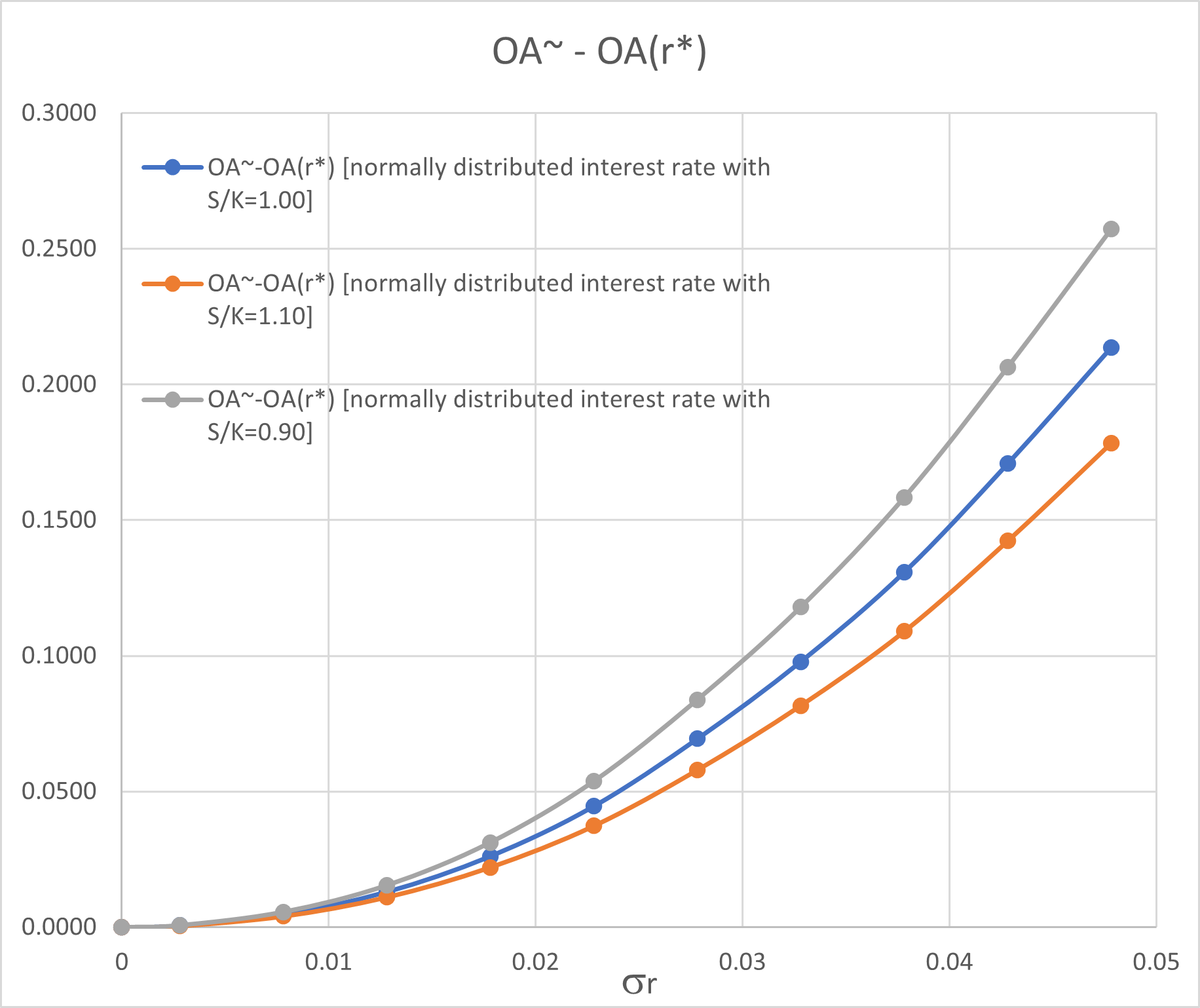}
  \caption{Optionality versus standard deviation for an equity put under a normally distributed local interest rate for different moneyness levels }\label{fig:moneyness}
  \end{figure}

\begin{figure} [h!]
 \centering
   
 \includegraphics[scale=0.5]{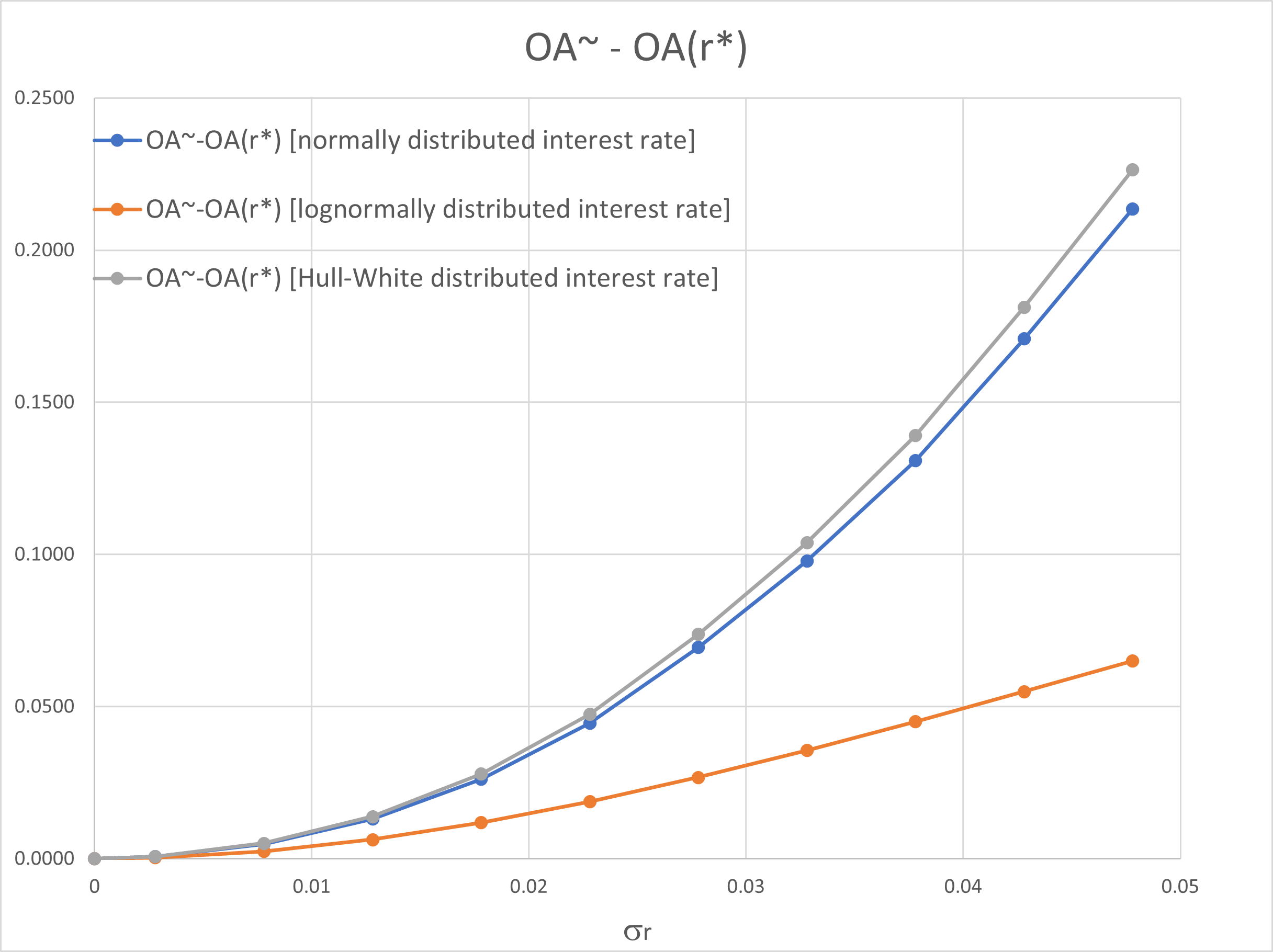}
  \caption{Optionality versus standard deviation for a an equity put option under different rate dynamics with $S/K=1.00$}\label{fig:distributions1}
  \end{figure}

In this section, we consider the American equity put with the same parameters as before. We begin by extending the normal interest rate framework through additional experiments, using two other fixed values of $S/K$, and we present the corresponding results in Figure \ref{fig:moneyness}. We observe that, the lower the moneyness levels, the greater is the optionality. 
Then, we consider a stochastic rate that is Hulll-White and lognormally distributed, with the results reported in 
Figure \ref{fig:distributions1}. We observe a behavior similar to that under a normal distribution, as the standard deviation $\sigma_r$ increases, the level of hidden optionality becomes more pronounced. Again, the results show consistent monotone optionality values, $\pi_A$, as a function of the interest rate volatility.

\subsubsection{Currency Put}\label{BIII2}
\begin{figure} [h!]
 \centering
   
 \includegraphics[scale=0.4]{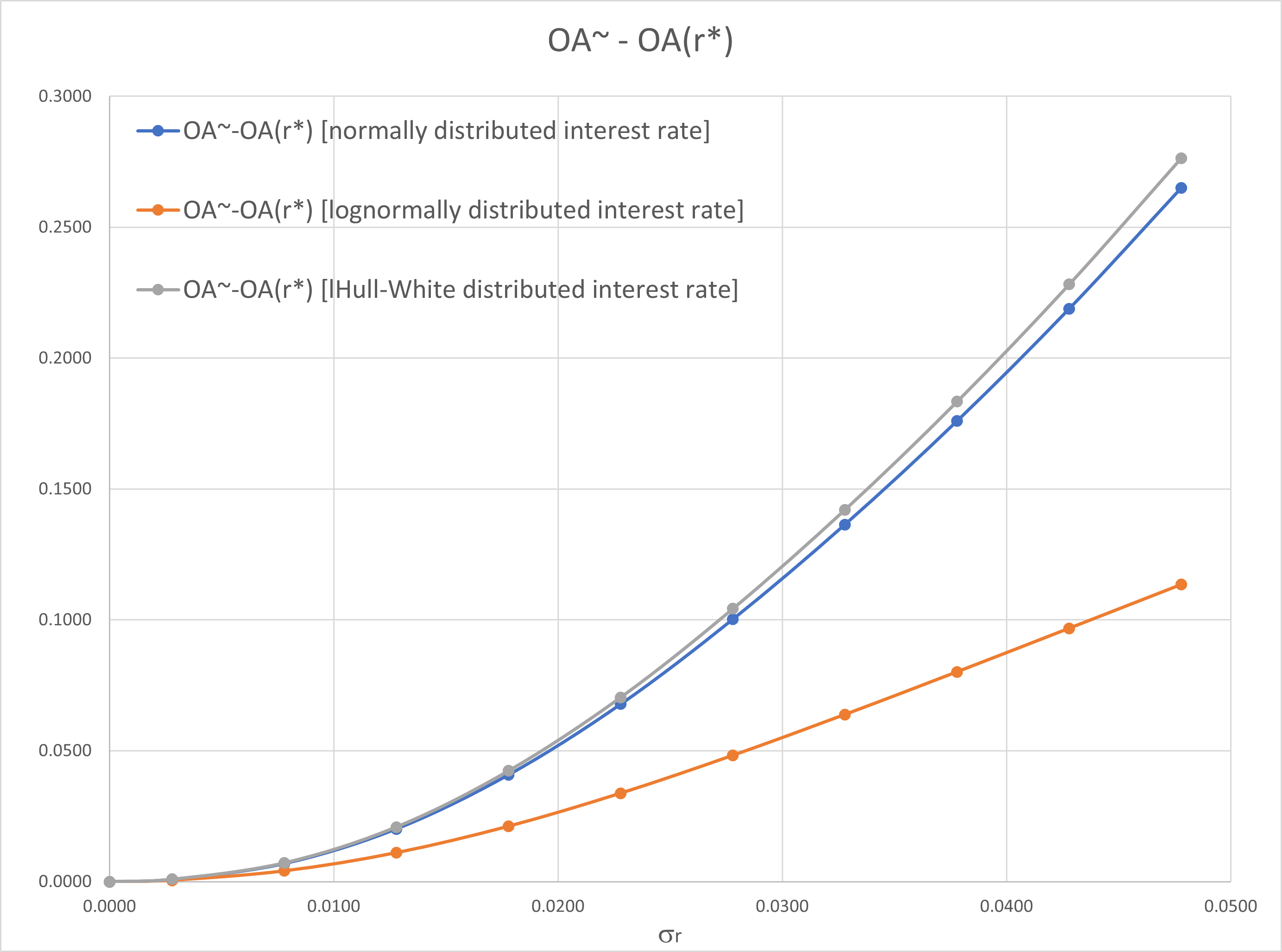}
  \caption{Optionality versus standard deviation for currency put option under different rate dynamics with $S/K=1.00$}\label{fig:distributions2}
  \end{figure}

\begin{figure} [h!]
 \centering
   
 \includegraphics[scale=0.5]{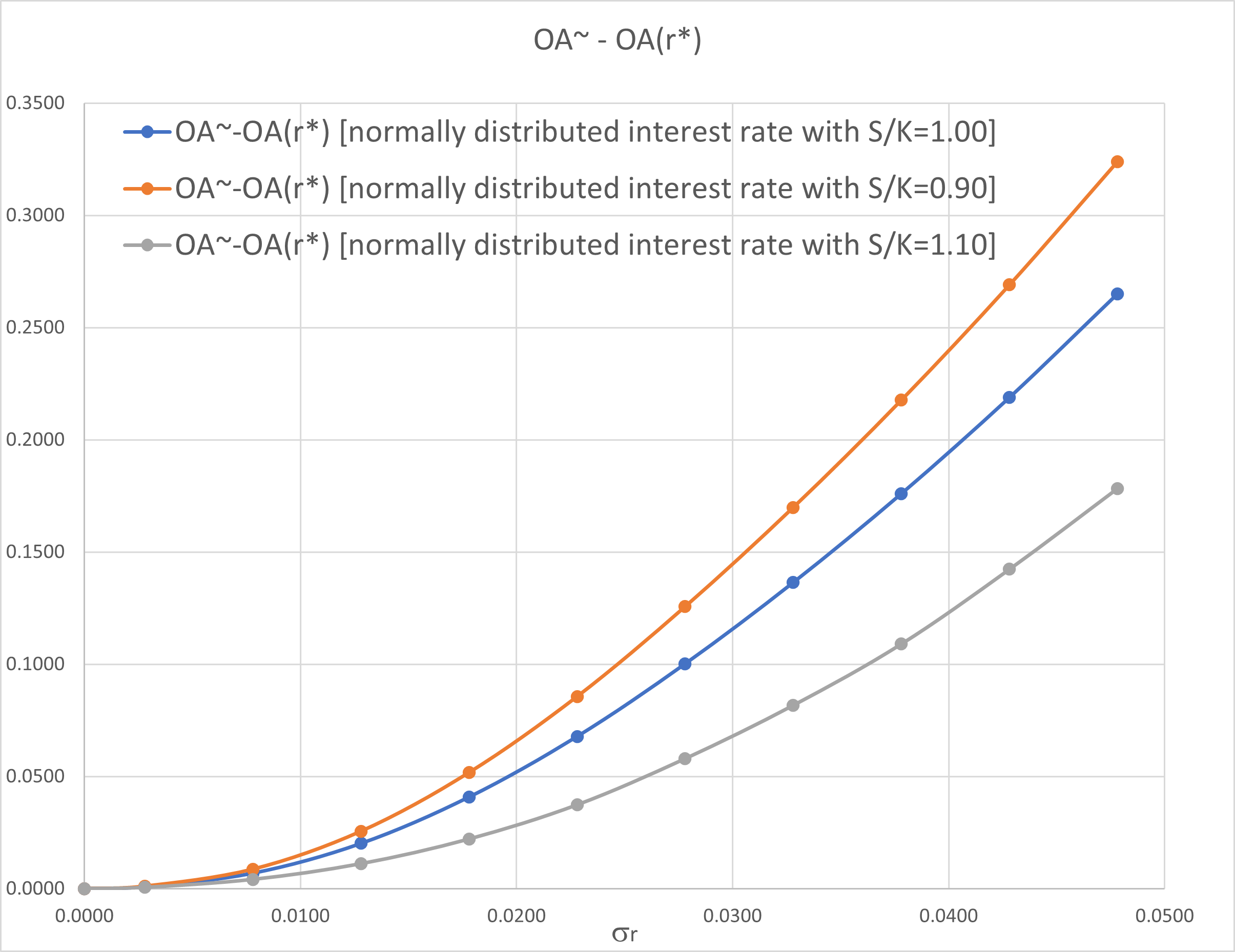}
  \caption{Optionality versus standard deviation for a currency put under a normally distributed local interest rate for different moneyness levels }\label{fig:moneyness2}
  \end{figure}

In this section, we further study the optionality of the American currency put, with the base parameters listed in Table \ref{tab:inputparameterscurput}. For $S/K=1$ and under a local rate, $r_1$ which is (i) normally, (ii) log-normally and (iii) Hull-White distributed, the optionality metric $\pi_A$ is computed similar to the case equity puts in section \ref{sec:numexperiments}. We vary again the standard deviation of the interest rate $\sigma_r$ and evaluate the resulting $\pi_A$, which increases with $\sigma_r$, as illustrated in 
Figure \ref{fig:distributions2}.  Moreover, we compute the optionality measure $\pi_A$ for three different moneyness levels and we present the results in Figure \ref{fig:moneyness2}. The results confirm once again that the deeper the option is in the money, the greater is the optionality.

\vspace{-1 mm}
\subsubsection{Currency Call}\label{BIII3}
\begin{figure} [h!]
 \centering
   
 \includegraphics[scale=0.6]{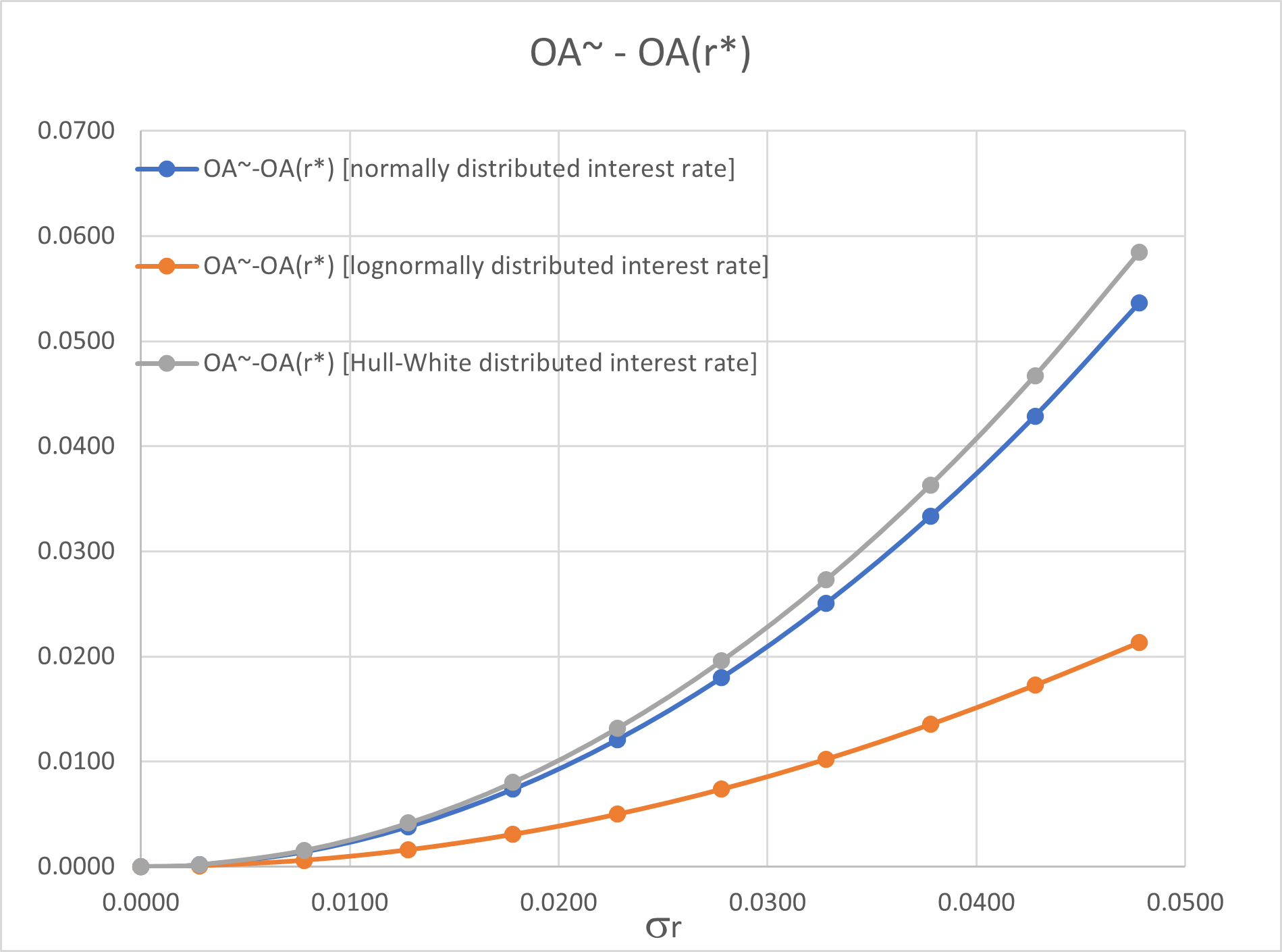}
  \caption{Optionality versus standard deviation for a currency put option under different rate dynamics with $S/K=1.00$}\label{fig:distributions3}
  \end{figure}

\begin{figure} [h!]
 \centering
   
 \includegraphics[scale=0.6]{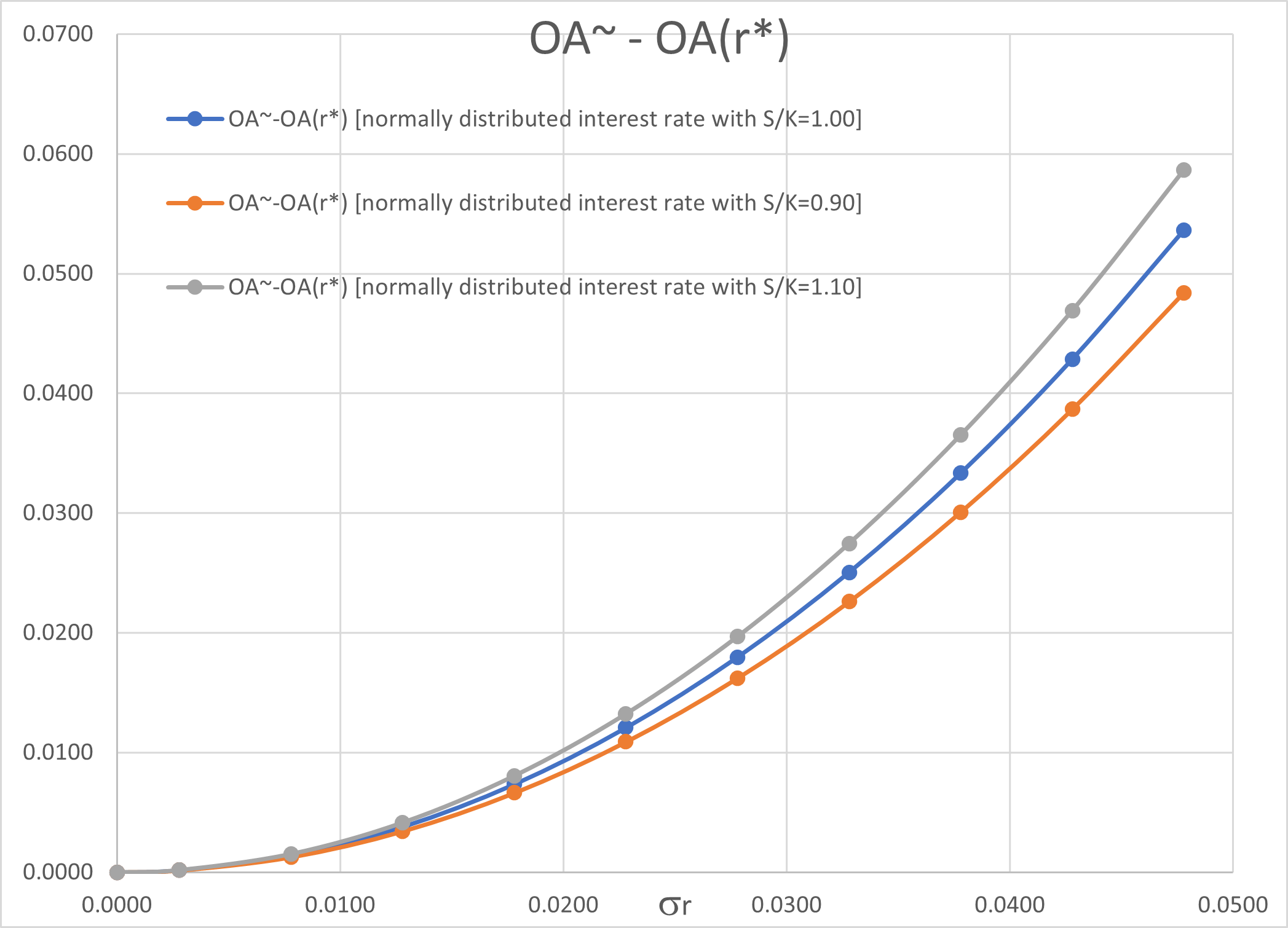}
  \caption{Optionality versus standard deviation for a currency call under a normally distributed local interest rate for different moneyness levels }\label{fig:moneyness3}
  \end{figure}

Next, we explore  currency call options using similar parameters to those in Section \ref{BII3}, and under various local rate distributions. The results are shown in 
and Figure \ref{fig:distributions3} confirm a consistent monotonic behavior of the optionality metric as a function of the rate volatility. Then, we consider three different moneyness levels and repeat the same experiments. The results in Figure \ref{fig:moneyness3} reveal again that deep in the money currency calls (with high values of the moneyness $S/K$) exhibit higher optionality.

\subsection{ Optionality of American vs European}\label{BIV}
\begin{figure} [h!]
 \centering
   
 \includegraphics[scale=0.6]{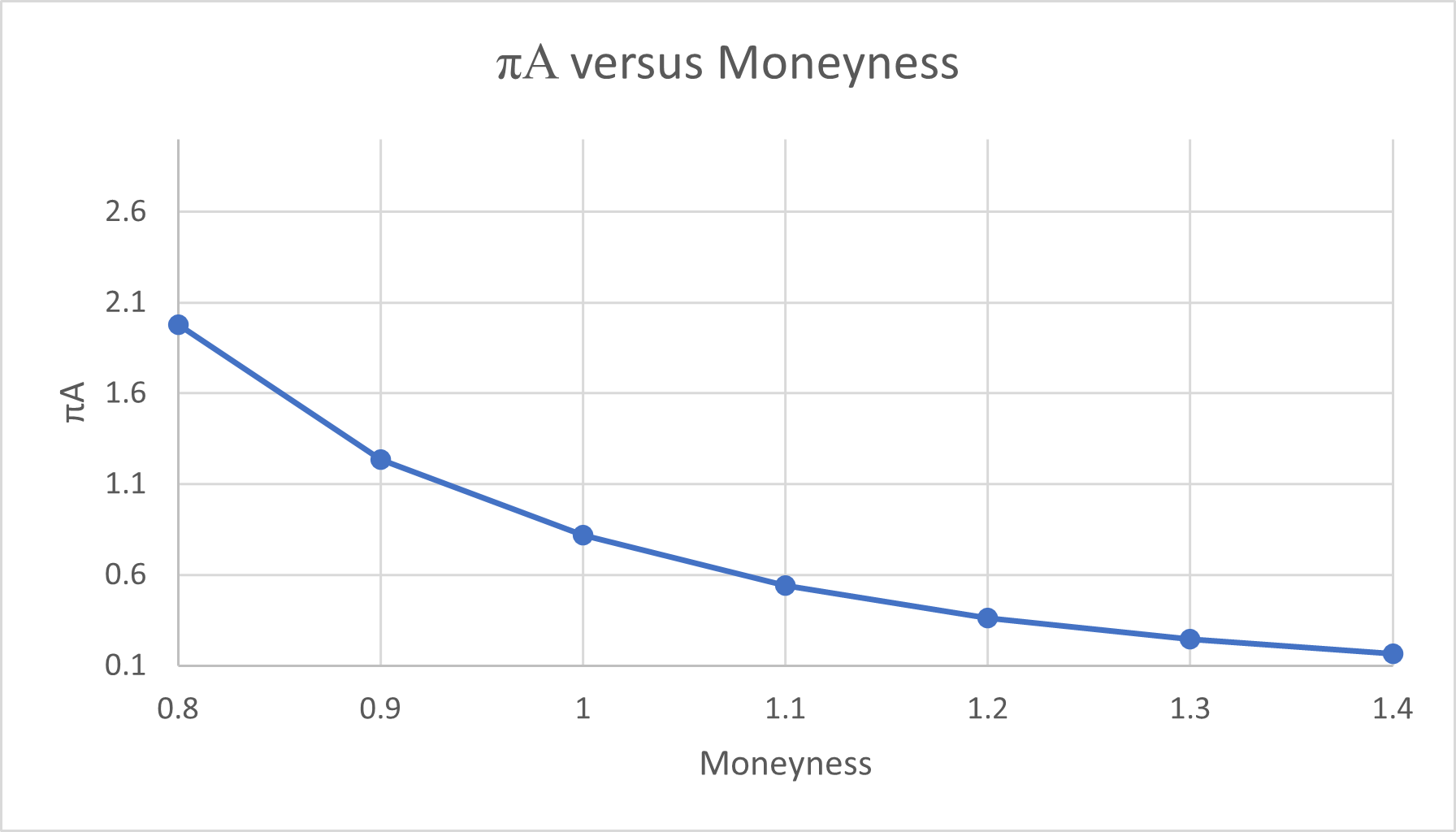}
  \caption{$\pi_A^{(2)}
  $ versus moneyness for equity put options under a normally distributed interest rate}\label{fig:diffequityputs}
  \end{figure}

\begin{figure} [h!]
 \centering
   
 \includegraphics[scale=0.6]{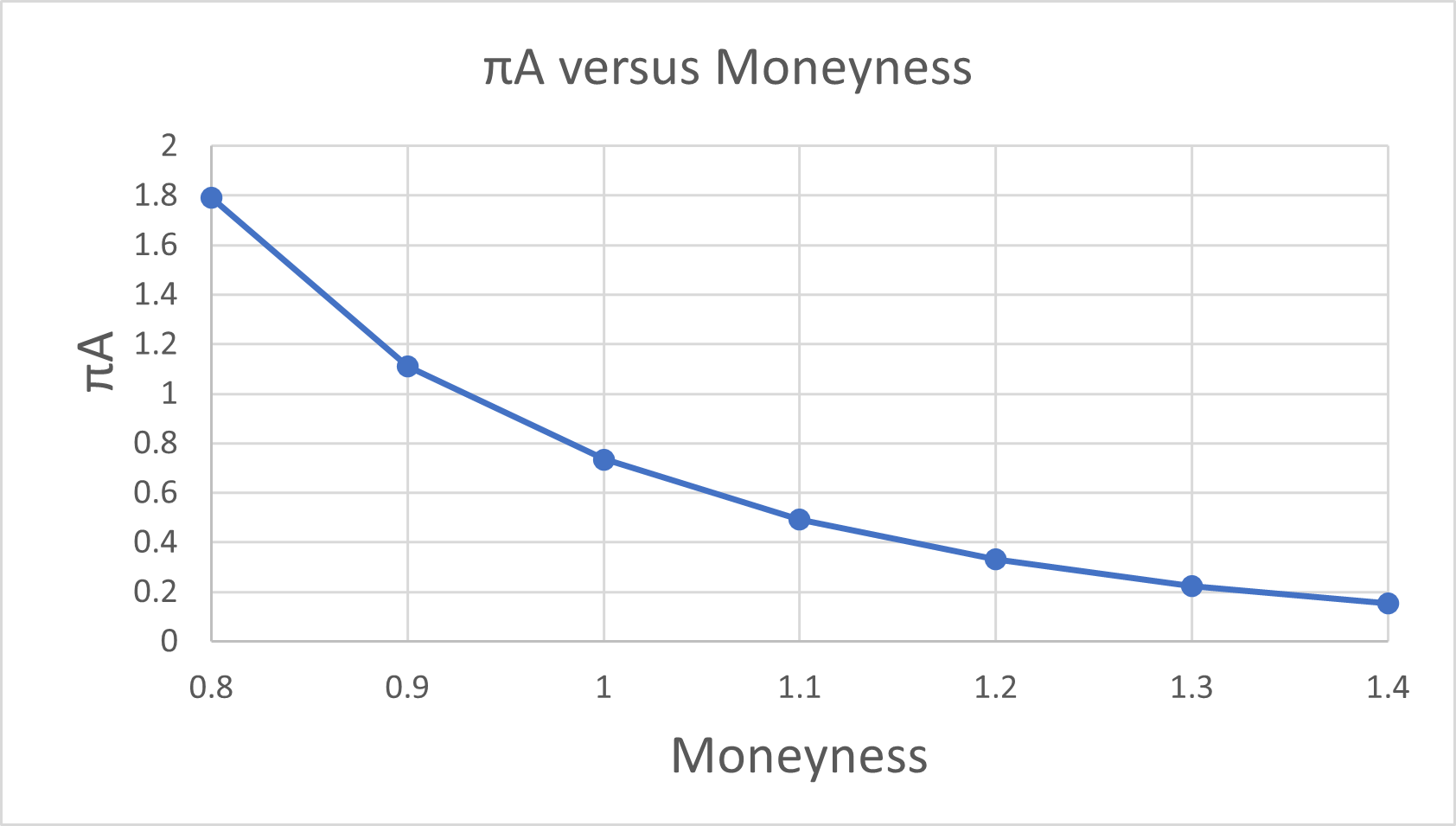}
  \caption{$\pi_A^{(2)}$ versus moneyness for currency put options under a normally distributed interest rate}\label{fig:diffcurrencyyputs}
  \end{figure}

\begin{figure} [h!]
 \centering
   
 \includegraphics[scale=0.6]{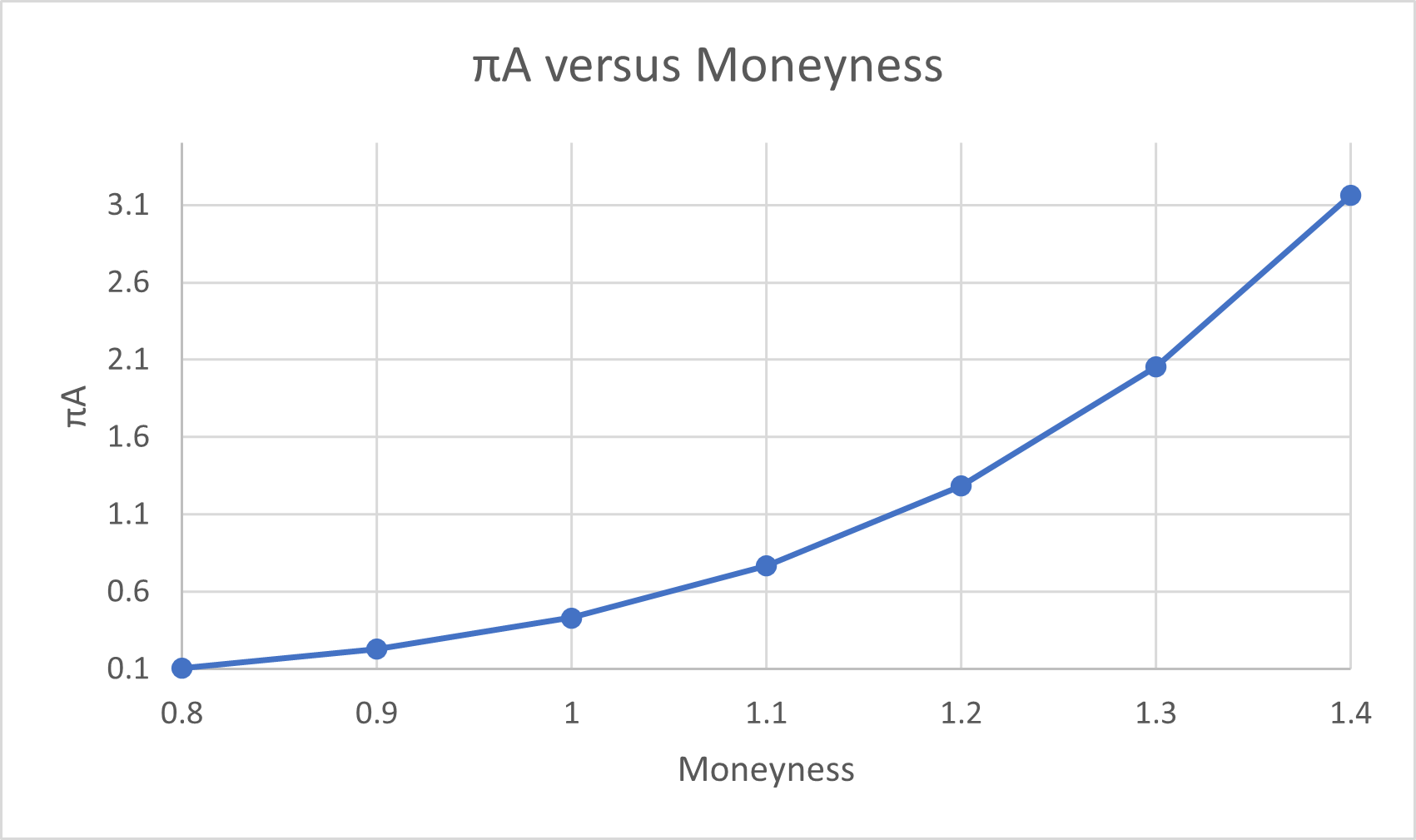}
  \caption{$\pi_A^{(2)}$ versus moneyness for currency call options under a normally distributed local interest rate}\label{fig:diffcurrencycalls}
  \end{figure}

As an alternative measure of hidden optionality, we can compute the stochasticized American option value, $\widetilde{O}A(r)$, under a normal interest rate, and compare it with the corresponding European price, estimated as 
\[
\widetilde{O}E(r)=\int_{D_r} OE(r)f_{r_T}(r)dr,
\]
where $f_{r_T}(r)$ is the density of the interest rate at the option maturity, $T$. We then estimate 
\[
\pi_A^{(2)}=\widetilde{O}A(r)-\widetilde{O}E(r), 
\]
as a measure of the gain from the hidden optionality of the American option. We compute $\pi_A^{(2)}$ for various moneyness levels for (i) an equity put, (ii) a currency put, and (iii) a currency call. We present the results in Figures \ref{fig:diffequityputs}-\ref{fig:diffcurrencycalls}.
The results show again consistent positive $\pi_A^{(2)}$ values, indicating that the American option is less vulnerable to interest rate stochasticity than the European one. In addition, we observe that the lower the moneyness level, the greater is the optionality (and implicitly the robustness) of the American option over the European one. Such results have been observed, sporadically, in the literature, e.g., \cite{GarmanKoh} and \cite{MedvedevScaillet2010}.
\section{Discussion}
This paper uncovered significant hidden convexity in American options and presented a technique for pricing and uncovering hidden risks so far ignored in the literature. While our approach was near-exhaustive, by focusing on one single interest rate, even simpler techniques can be applied on the fly, see Eq. \ref{simplifiedeq} as $\pi_{\Delta a}$ could be immediately computed by moving either rate.


\begin{thebibliography}{999}
\bibitem{taleb2013mathematical}
Taleb, N.N. and Douady, R. (2013). Mathematical definition, mapping, and detection of (anti) fragility.\textit{Quantitative Finance}.
\bibitem{taleb2018IMF}
Taleb, N.N., Canetti, E., Kinda, T., Loukoianova, E., and Schmieder, C.
  (2018).
\newblock A new heuristic measure of fragility and tail risks: application to
  stress testing.
\newblock {\em International Monetary Fund}.
\bibitem{Bachelier1900}
Bachelier L. Th\'eorie de la sp\'eculation. \textit{Ann Sci \'Ec Norm Sup\'er}. 1900;17:21--86.
\bibitem{Dupire1994}
Dupire B. Pricing with a smile. \textit{Risk}. 1994;7(1):18--20.
\bibitem{Dupire1997}
Dupire B. A unified theory of volatility. In: \textit{Derivatives pricing and credit exposures}. London: Risk Publications; 1997.
\bibitem{DermanKani1994}
Derman E, Kani I. The volatility smile and its implied tree. \textit{Quantitative Strategies Research Notes}. New York (NY): Goldman Sachs; 1994.
\bibitem{HullWhite1987} 
Hull J, White A. The pricing of options on assets with stochastic volatilities. \textit{J Finance}. 1987;42(2):281--300.
\bibitem{Gatheral2006}
Gatheral J. The volatility surface: a practitioner's guide. Hoboken (NJ): John Wiley \& Sons; 2006.


\bibitem{Taleb1997}
Taleb N. N. Dynamic hedging: managing vanilla and exotic options. New York (NY): John Wiley \& Sons; 1997.
\bibitem {HullBook}
Hull JC. Options, futures, and other derivatives. 10th ed. Harlow: Pearson; 2017.
\bibitem{GarmanKohlhagen1983}
Garman MB, Kohlhagen SW. Foreign currency option values. \textit{J Int Money Finance}. 1983;2(3):231--7.
\bibitem {MedvedevScaillet2010}
Medvedev A, Scaillet O. Pricing American options under stochastic volatility and stochastic interest rates. \textit{J Financ Econ}. 2010;98(1):145--68.
\bibitem {StoerBulirsch2013}
Stoer J, Bulirsch R. Introduction to numerical analysis. 3rd ed. New York (NY): Springer Science \& Business Media; 2013 Mar 9. Chapter 3, Topics in integration.
\bibitem {FRED2025}
FRED, Federal Reserve Bank of St. Louis, 2025. Market yield on U.S. Treasury securities at 1-year constant maturity, quoted on an investment basis [DGS1]. 
Available at: \texttt{https://fred.stlouisfed.org/series/DGS1}

\bibitem {ElHassanandMaddah}
Hassan NE, Maddah B. Power approximation for pricing American options, 2026. \textit{Int Trans Oper Res}. 2024 Jan;33(1), :117--1-42.

\bibitem {Cox}
Cox, J. C., Ross, S. A. and Rubinstein, M., 1979. Option pricing: a simplified approach. 
\textit{Journal of Financial Economics}, 7(3), pp.229--263.
\bibitem {Black}
Black, F. and Scholes, M., 1973. The pricing of options and corporate liabilities. 
\textit{Journal of Political Economy}, 81(3), pp.637--654.
\bibitem{GarmanKoh}
Garman, M. B. and Kohlhagen, S. W., 1983. Foreign currency option values. 
\textit{Journal of International Money and Finance}, 2(3), pp.231--237.
\end{thebibliography}
\end{document}